\documentclass[letterpaper,twocolumn,10pt]{article}
\usepackage{usenix}

\usepackage{amsmath}
\usepackage{amssymb}
\usepackage{booktabs}
\usepackage{float}
\usepackage{graphicx}
\usepackage{pifont}
\usepackage{listings}
\usepackage[ruled,vlined,linesnumbered,noend,nosemicolon]{algorithm2e}
\usepackage{tikz}
\usetikzlibrary{arrows.meta,positioning,fit,calc,backgrounds}
\pgfdeclarelayer{background}
\pgfsetlayers{background,main}
\usepackage{xspace}
\newcommand{\system}{\textsc{Portico}\xspace}

\providecommand{\Description}[1]{}
\definecolor{snippetbg}{RGB}{248,248,248}
\definecolor{snippetframe}{RGB}{196,196,196}
\lstdefinestyle{porticosnippet}{
  basicstyle=\ttfamily\scriptsize,
  columns=fullflexible,
  keepspaces=true,
  breaklines=true,
  breakatwhitespace=false,
  showstringspaces=false,
  frame=single,
  framerule=0.3pt,
  rulecolor=\color{snippetframe},
  backgroundcolor=\color{snippetbg},
  xleftmargin=0.35em,
  xrightmargin=0.35em,
  aboveskip=0.35em,
  belowskip=0.15em
}

\title{Lingering Authority: Revocable Resource-and-Effect Capabilities \\for Coding Agents}

\author{\rm{Igor Santos-Grueiro}\\ International University of La Rioja}

\begin{document}

\date{}
\maketitle

\begin{abstract}
Coding agents often receive broad tool access for an entire task, even when a
resource is needed only for one subgoal. We call this gap
\textit{lingering authority}: a temporary resource/effect capability remains
exposed after the episode that justified it has closed.

\system is a reference monitor for revocable capabilities exposed to the
planner. It compiles an explicit task contract into initial capabilities, grant
rules, trusted closure predicates, and global deny rules. A
request--grant--invoke lifecycle materializes expansions as opaque,
epoch-bound handles. Closure removes those handles from the next planner
interface and rejects stale replay before side effects. The monitor assumes
mediated tools and a sound typed catalog.

In controlled coding-agent tasks, \system records no executed contract-forbidden
effects in the evaluated runs, while controlled grants recover boundary work
blocked by a fixed narrow envelope. A non-revoking comparator
receives the same initial envelope and the same grants at the same turns. On
the closure slice, both systems match task success, scope compliance, and all
pre-closure decisions; \system then rejects 10/10 post-closure reuses, while
the comparator permits 10/10. A deterministic stale-write audit records 0/6
versus 6/6 executed forbidden effects. Scripted traces and six live model
traces over file writes, git mutation, and network egress show the same split.
In a four-episode same-policy diagnostic, broad request exposure preserves
zero executed forbidden effects but raises blocked proposals from 67 to 84.
Frozen real-repository runs, with commits and traces recorded,
exercise the same lifecycle on real project layouts.

\end{abstract}

\section{Introduction}

Coding agents turn natural-language tasks into tool calls over repositories,
tests, shells, package managers, network clients, and version control. They
often start with more authority than the task needs. The key problem is
temporal: authority justified for one subgoal can become stale once that subgoal
closes. If the agent later uses this stale authority, the planner-tool interface
has exposed \emph{lingering authority}: a temporary resource/effect capability
that remains exposed after the grant episode that justified it closed.

Consider a task that propagates a timeout from \path{src/api.py} to
\path{src/serialization.py}. The initial phase needs the API file, the focused
API test, and the test command. The contract declares serialization access as
an on-demand boundary capability. A planner request may select that declared
rule, but it cannot create new authority. A plausible note such as
\path{notes/packaging.md} is outside the task. The desired behavior is to deny
the note, grant serialization authority for the boundary episode, and revoke
that authority when the episode closes. A static allowlist or sandbox may make
the serialization file reachable for the whole run. The security question is
whether it should remain available to the planner for the whole run.

Existing defenses address adjacent layers: hostile content or proposed
actions~\cite{promptinjection,injecagent,asb,toolemu,agentdojo,agentharm,agentalign,toolhijacker,promptarmor,llamafirewall},
tool and protocol trust~\cite{mcp_landscape,breakingprotocol,securingmcp,smcp,mcpsecbench,mcptox,systematicmcp,mcpitp,ama,maltool},
and reachability or execution checks after tool selection~\cite{isolategpt,sandboxescapebench,agentcgroup,ace,saga,agentspec,pcas,maris,miniscope,agentbound,progent,cmtf,racg}.
They leave an earlier choice open: which otherwise admissible authority should be
visible to the planner now, and when should temporary authority disappear? If a
broad read option remains in the planner interface, the model can keep
planning around it even when a later monitor would reject some calls. The
exposed interface is therefore part of the security state.

We use \emph{capabilities exposed to the planner} for resource- and
effect-specific authority exposed through the mediated interface as current
handle state. They are distinct from runtime availability, sandbox
reachability, and execution checks. Revocation leaves prior context
intact: a resource name may still appear in the conversation, but authority is
absent when no mediated invocation over that resource can be selected or
validated without a fresh grant.

This paper studies bounded capability lifetime under fixed host isolation
and explicit task contracts. \emph{Exposure minimality} keeps unjustified
capabilities out of the interface. \emph{Justified expansion} permits
legitimate boundary work through explicit grants. \emph{Authority closure}
removes temporary authority when the reason for exposing it ends.

We implement these obligations in \system, a reference monitor for coding
agents. Given a task contract and typed tool catalog, \system compiles an
initial envelope, grant rules, closure rules, and global deny rules. \system
core exposes only the initial envelope. \system full adds the
request--grant--invoke lifecycle with trusted closure. A grant mints
epoch-bound handles for the covered resource and effect; closure removes those
handles from the next planner interface and rejects replay at execution time.
\system complements sandboxing and content defenses by making temporary
capabilities explicit, bounded, and auditable.

The evaluation uses three controlled fixture suites and one pinned
real-repository suite to test exposure, expansion, closure, and real-layout
behavior. A
non-revoking comparator isolates closure under the same grants, an all-visible
same-policy comparator isolates interface breadth under the same execution
policy, and a coarse action-filter diagnostic tests broad adjacent authority.
The study is relative to the written contract: given an explicit task contract, mediated tool
interface, and typed invocation-to-authority mapping, we compare upfront
exposure, execution-time denial, and bounded grant episodes with closure.

This paper makes three contributions:
\begin{itemize}
    \item It defines no-stale-use for resource- and effect-specific
    capabilities exposed to the planner: authority from a closed grant must
    disappear from later interfaces and be invalid at execution.
    \item It formalizes exposure minimality, justified expansion, and authority
    closure around a request--grant--invoke protocol, and implements them in
    \system with epoch-bound grant handles.
    \item It evaluates bounded lifetime on controlled fixtures and pinned real
    repositories, showing reduced post-closure stale-capability availability and
    lower blocked-proposal pressure than an all-visible same-policy interface
    in the post-closure diagnostic.
\end{itemize}

\paragraph*{Artifact Availability.} An anonymized artifact repository for this
submission is available at
\url{https://anonymous.4open.science/r/portico/}. It contains the checked-in
task contracts, frozen result roots, reproduction scripts, public
\texttt{portico-tool} traces, and secondary \system decision core used to audit
the paper.

\section{A Concrete Authority-Lifecycle Trace}
\label{sec:motivating-trace}

The running timeout task exposes lingering authority with one false side
request and one legitimate boundary request. The task is to propagate an
optional timeout parameter from
\path{src/api.py} into \path{src/serialization.py}. At the start, the planner
only needs the API file, the focused API test, and pytest. The serialization
file is declared as an on-demand boundary capability; a planner request can
select that existing rule for evaluation but cannot create the rule. Packaging
notes such as \path{notes/packaging.md} are outside the task.
Table~\ref{tab:main-lifecycle-trace} gives the step-by-step trace.

\begin{table}[h]
\centering
\caption{Timeout-propagation authority trace.}
\label{tab:main-lifecycle-trace}
\scriptsize
\setlength{\tabcolsep}{3pt}
\begin{tabular}{p{0.08\columnwidth}p{0.39\columnwidth}p{0.43\columnwidth}}
\toprule
\textbf{Step} & \textbf{Planner event} & \textbf{\system decision and visible capabilities} \\
\midrule
0 & Start task. & Expose $E_0$: \path{src/api.py}, \path{tests/test_api.py}, and focused pytest. \\
1 & Read \path{notes/packaging.md}. & Deny; packaging notes are never justified. Visible capabilities unchanged. \\
2 & Request serialization authority. & Grant serialization capability; mint epoch-bound handles for the next planner interface. \\
3a & Read \path{src/serialization.py} through the read handle. & Permit under the live grant handle. \\
3b & Edit \path{src/serialization.py} through the write handle. & Permit under the live grant handle. \\
4 & Focused tests pass. & Revoke the serialization grant; visible capabilities return to the initial API scope. \\
5a & Replay the serialization handle. & Deny; the handle is stale after closure. \\
5b & Reread \path{notes/packaging.md}. & Deny; packaging notes were never in the contract. \\
\bottomrule
\end{tabular}
\end{table}

\path{src/serialization.py} changes status over time. It is unjustified before
the boundary, justified during the boundary, and stale after the boundary
closes. The grant request does not execute the file operation by itself; it
materializes handles for the next planner step. If those handles remain
exposed after closure, they have become lingering authority. The
following sections unpack this trace as a contract, typed tools, an envelope, a
grant, a lifetime, a revoke event, a trace, and a metric.

The same pattern appears in temporary-read episodes. A renderer repair can
start with \path{src/render.py}, \path{tests/test_build.py}, and pytest
visible. During inspection, a recheck note may be justified as read-only
context; after the inspection subgoal closes, the note should disappear from
the planner interface while local source and test authority remain. This
second shape shows the same separation: legitimate expansion should be
possible, and temporary authority should expire when its subgoal ends.

\section{Problem and Threat Model}
\label{sec:threat}

We study authority exposure before a tool call runs. The planner is untrusted:
it proposes tool invocations, and the runtime arbiter decides which invocations
exist on the mediated interface and which ones may execute. Host isolation is
held fixed. Our question is which task capabilities should be exposed
now, and when temporary authority should disappear. The running example and
Figure~\ref{fig:authority-surfaces} illustrate the distinction: reachability,
execution-time denial, plan integrity, and task-wide legitimacy still leave open
which capabilities are exposed at the current phase. With host and sandbox
fixed, planner outputs are untrusted requests, and the only variable is the
capability interface exposed through the mediated path before execution.

\paragraph{System model.}
We consider a coding agent whose planner emits structured invocations over a
repository and a small set of external resources. The tool interface includes
file reads and writes, shell/test execution, dependency management, network
fetch, and git-related tooling. A runtime authority arbiter sits between the
planner and those tools. Every compared condition uses the same execution
substrate, sandbox profile, and low-level mediation path; what changes is the
authority exposed to the planner before execution.

\paragraph{Capabilities exposed to the planner.}
A capability is exposed to the planner when it is present on the mediated interface:
tool schemas, target scopes, argument constraints, active grant options, and
deny or escalation responses that can affect later planning. We distinguish a
request option, which lets the planner formulate a request, from an
executable capability, which can produce effects after validation. Authority
can remain available in the runtime without being exposed to the planner. Revocation
leaves context intact: a path name may remain in the conversation. Effect
authority is absent when no mediated invocation over that resource can be
successfully validated without a fresh grant.

\paragraph{Adversary capabilities.}
The adversary influences the planner through ordinary inputs: repository
files, issue text, retrieved documentation, tool descriptors or protocol
metadata, and misleading tool outputs. We include indirect prompt injection,
misleading metadata, dependency mutation in dependency-stable tasks,
read-plus-network pivots, and over-broad git or package-manager authority. A
legitimate tool with surplus exposed capability is enough; runtime compromise
or bypass of the mediated path is outside the threat model.

\paragraph{Security goal.}
The goal is to expose only task and phase justified authority to the planner,
deny out-of-contract invocations before execution, and revoke temporary
authority after its justification ends. This limits avoidable interface risk
while lower-level isolation continues to enforce runtime boundaries. Planner
outputs are requests to adjudicate, not declarations of need.

\paragraph{Trusted base and non-goals.}
The trusted base includes the policy engine, typed tool catalog, enforcement
hooks, grant/revoke state, and host runtime or external sandbox. Process
isolation, syscall filtering, browser sandboxing, and network mediation remain
complementary. We exclude compromised kernels, sandbox bypasses, covert
channels, complete protocol coverage for every tool ecosystem, tool
implementations that bypass the invocation interface, unstructured tool use
outside the mediated call path, and optimal contract compilation. The next
section separates availability, reachability, admissibility, and
executable capability.

\section{Model of Capabilities and Interfaces}
\label{sec:formalism}

\paragraph{Planner interface as runtime state.}
For task $\tau$, the monitor decides which tool-mediated actions can be
proposed through the planner interface and which capabilities can
produce effects after validation. Each requested action maps through the typed
catalog to an authority descriptor
\[
\beta=(u,p,r,e,\iota,\phi),
\]
where $u$ is the tool primitive, $p$ the primitive privilege, $r$ the resource,
$e$ the side-effect class, $\iota$ the task intent, and $\phi$ the trusted phase.
This per-request mapping matters because one tool, especially shell, can induce
many resources and effects~\cite{saltzer,seccomp,capsicum,cheri}.

\begin{figure}[t]
\centering
\begin{tikzpicture}[
  font=\scriptsize\sffamily,
  layer/.style={draw, rounded corners=6pt, thick},
  panel/.style={draw, rounded corners=6pt, thick},
  surfacelabel/.style={
    font=\scriptsize\sffamily\bfseries,
    inner sep=1.5pt,
    rounded corners=1pt,
    align=center
  },
  rowlabel/.style={
    anchor=west,
    align=left,
    text width=1.48cm,
    font=\scriptsize\sffamily
  },
  rowvalue/.style={
    anchor=east,
    align=right,
    text width=1.10cm,
    font=\scriptsize\sffamily
  },
]
\node[font=\scriptsize\sffamily\bfseries, anchor=west] at (-4.05,2.9) {Sets};
\node[font=\scriptsize\sffamily\bfseries, anchor=west] at (0.72,2.9) {Example};

\node[layer, fill=gray!10, draw=gray!55,
      minimum width=4.55cm, minimum height=5.55cm] (available) at (-2.05,0) {};
\node[layer, fill=blue!7, draw=blue!50!black,
      minimum width=3.50cm, minimum height=3.70cm] (reachable) at (available.center) {};
\node[layer, fill=green!8, draw=green!50!black,
      minimum width=2.65cm, minimum height=2.60cm] (admissible) at (available.center) {};
\node[layer, fill=orange!12, draw=orange!60!black, line width=1.2pt,
      minimum width=1.75cm, minimum height=1.05cm] (visible) at (available.center) {};

\node[surfacelabel, anchor=north west]
  at ([xshift=8pt,yshift=-10pt]available.north west) {Available};

\node[surfacelabel, anchor=north east]
  at ([xshift=-8pt,yshift=-5pt]reachable.north east) {Reachable};

\node[surfacelabel, anchor=south west]
  at ([xshift=8pt,yshift=6pt]admissible.south west) {Policy-admissible};

\node[surfacelabel, text width=1.55cm]
  at (visible.center) {Executable\\capability};

\node[panel, minimum width=3.55cm, minimum height=5.25cm] (example) at (2.25,0) {};

\node[align=center, text width=3.05cm, font=\scriptsize\sffamily]
  at (example.north) [yshift=-0.55cm]
  {\texttt{read src/serialization.py}\\during timeout propagation};

\draw[black!25]
  ([xshift=0.22cm,yshift=-1.1cm]example.north west) --
  ([xshift=-0.22cm,yshift=-1.1cm]example.north east);

\fill[gray!35]
  ([xshift=0.24cm,yshift=-1.58cm]example.north west) rectangle ++(0.12,0.12);
\node[rowlabel]
  at ([xshift=0.46cm,yshift=-1.52cm]example.north west) {\textbf{Available}};
\node[rowvalue]
  at ([xshift=-0.26cm,yshift=-1.52cm]example.north east) {\textbf{yes}};

\fill[blue!40]
  ([xshift=0.24cm,yshift=-2.28cm]example.north west) rectangle ++(0.12,0.12);
\node[rowlabel]
  at ([xshift=0.46cm,yshift=-2.22cm]example.north west) {\textbf{Reachable}};
\node[rowvalue]
  at ([xshift=-0.26cm,yshift=-2.22cm]example.north east) {\textbf{yes}};

\fill[green!45!black]
  ([xshift=0.24cm,yshift=-2.98cm]example.north west) rectangle ++(0.12,0.12);
\node[rowlabel]
  at ([xshift=0.46cm,yshift=-2.92cm]example.north west) {\textbf{Policy/}\\\textbf{grant eligible}};
\node[rowvalue]
  at ([xshift=-0.26cm,yshift=-2.92cm]example.north east) {\textbf{yes}};

\fill[orange!65!black]
  ([xshift=0.24cm,yshift=-3.82cm]example.north west) rectangle ++(0.12,0.12);
\node[rowlabel]
  at ([xshift=0.46cm,yshift=-3.76cm]example.north west) {\textbf{Executable}\\\textbf{now}};
\node[rowvalue]
  at ([xshift=-0.26cm,yshift=-3.76cm]example.north east) {\textbf{no}};

\node[align=center, text width=7.55cm, font=\scriptsize\sffamily\itshape]
  at (0,-3.18)
  {$A_{\mathit{avail},t} \supseteq A_{\mathit{reach},t} \supseteq A_{\mathit{adm},t} \supseteq V_{\beta,t}$};

\end{tikzpicture}
\caption{Four nested but non-equivalent descriptor sets. In the mediated configuration studied here, the innermost descriptor set is the projection of executable capabilities. Request options are modeled separately as part of the planner interface. Right: a secondary-module read may be available, reachable, and admissible under a future grant rule while still not being executable now.}
\Description{A two-panel diagram. The left panel shows four nested rectangles labeled Available, Reachable, Policy-admissible, and Executable capability. The right panel shows an example, reading src slash serialization dot py during timeout propagation, with row-by-row values: available yes, reachable yes, policy slash grant eligible yes, and executable now no. A line below summarizes the subset relation among the four sets.}
\label{fig:authority-surfaces}
\end{figure}
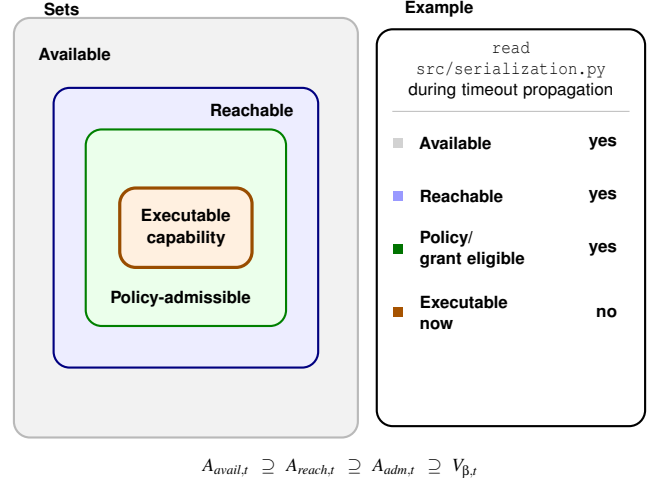

\paragraph{Four related but non-equivalent descriptor sets.}
Figure~\ref{fig:authority-surfaces} separates four sets over the same
descriptor type. Let $\mathcal{B}_\tau$ be the descriptor universe for task
$\tau$. $A_{\mathit{avail},t}$ contains descriptors in $\mathcal{B}_\tau$ whose
tool primitive is registered; $A_{\mathit{reach},t}$ restricts them to
runtime-reachable resources; $A_{\mathit{adm},t}$ keeps descriptors admissible
under policy or a future grant rule; and $V_{\beta,t}$ is the projection of
currently executable runtime capabilities back to descriptors. Request options
are interface state, but not effect authority until a request mints
a live handle.

\paragraph{Task- and phase-scoped state.}
For task $\tau$, the compiled policy starts from an initial authority envelope
\[
\mathcal{E}_0(\tau) = (T_0, \Pi_0, \mathcal{R}_0, C_0),
\]
where $T_0$ is the initial tool-schema set, $\Pi_0$ the initial primitive
privileges, $\mathcal{R}_0$ the initial resources, and $C_0$ the initial
contextual constraints. The monitor maintains phase, live grants, and audit
state, then derives executable capabilities from $\mathcal{E}_0$ plus live
grants.

An \emph{authority descriptor} is a concrete requested operation
\[
\beta = (u,p,r,e,\iota,\phi),
\]
where $u$ is the tool, $p$ the primitive privilege required by that invocation,
$r$ the resource or resource class, $e$ the side-effect class, $\iota$ the
task intent, and $\phi$ the phase under which the invocation is being
adjudicated. A \emph{runtime capability instance} is
$\alpha=(\beta,\kappa)$, where $\kappa$ records compiler-issued initial
authority or a grant ID, epoch, and opaque handle. Runtime capabilities live in
$\mathcal{B}_\tau\times K$ for handle metadata space $K$. At runtime, $\phi$ is
derived from $\phi_t$ and checked by \texttt{phase\_ok}; it remains in $\beta$
for audit.
For example, the temporary serialization read in the running trace maps to:
\[
\begin{array}{ll}
u=\texttt{read\_file}, & p=\texttt{read},\\
r=\texttt{src/serialization.py}, & e=\texttt{inspect\_code},\\
\iota=\texttt{propagate\_timeout}, & \phi=\texttt{patch},\\
\kappa=(\textit{grant id},\textit{epoch},\textit{read handle}). &
\end{array}
\]
The packaging-note probe has the same primitive privilege but a different
resource and intent, and no grant can mint a capability for it.

The initial envelope induces compiler-issued authority instances:
\[
\begin{aligned}
\mathsf{Init}(\mathcal{E}_0,\tau,s_t)=
\{\alpha=(\beta,\kappa) \mid{}&
u(\beta) \in T_0 \wedge p(\beta) \in \Pi_0 \wedge{}\\
& r(\beta) \in \mathcal{R}_0 \wedge{}\\
& C_0(e(\beta),\iota(\beta),\phi(\beta),\kappa,\tau,s_t)\}.
\end{aligned}
\]
Each field filters part of the concrete invocation. Live grants add runtime
instances:
\[
\mathsf{GrantCaps}(L_t,\tau,s_t)=
\{\alpha \mid \mathsf{live}(\alpha,L_t,\Lambda,s_t)\}.
\]
The executable runtime capabilities are derived at each turn:
\[
\begin{aligned}
\mathsf{Cap}_t &=
\mathsf{Init}(\mathcal{E}_0,\tau,s_t) \cup
\mathsf{GrantCaps}(L_t,\tau,s_t),\\
V_t &= \{\alpha\in\mathsf{Cap}_t \mid
J_{\mathsf{Portico}}(\alpha,s_t)\},\\
V_{\beta,t} &= \{\beta \mid \exists \kappa.\;(\beta,\kappa)\in V_t\}.
\end{aligned}
\]
Let
$\mathcal{A}(\tau)$ be the authority instances inducible for task $\tau$ over
the run. $Q_t$ is the set of visible expansion rules or authority-request
endpoints, and $F_t$ the action descriptors the planner interface allows
the model to formulate. The concrete interface is a tagged union:
\[
I_t=\operatorname{Exec}(V_t)\uplus\operatorname{Prop}(F_t)\uplus
\operatorname{Req}(Q_t).
\]
Executable effect authority still requires a capability in $V_t$; selecting a
request rule does not grant authority to read or write the covered resource.
The all-visible same-policy comparator broadens $F_t$ while keeping $V_t$,
grant timing, closure events, and execution-time validation fixed.
\paragraph{Justification and grants.}
Here $J(\alpha,s_t)$ is the ideal task-justification relation for whether
runtime capability $\alpha$ should be executable in state $s_t$. \system
enforces the computable relation $J_{\mathsf{Portico}}$ compiled from the task
contract and typed catalog.
Let the compiled policy be
\[
P=(\mathcal{E}_0,\Gamma,\Lambda,G),
\]
where $\mathcal{E}_0$ is the initial envelope, $\Gamma$ is the set of declared
expansion rules, $\Lambda$ is the set of closure rules, and $G$ is the set of
global invariants. Let the runtime state be
\[
s_t=(\phi_t,L_t,O_t),
\]
where $\phi_t$ is the current phase, $L_t$ live grants, and $O_t$ observable
audit/run state. The sets $V_t$, $F_t$, and $Q_t$ are derived before each
planner turn.
\system treats planner requests as evidence; planner text alone cannot
authorize an action. In particular, the intent component $\iota$ is checked
through the typed catalog, contract selectors, phase, and observable state; a
planner-provided rationale is never enough to satisfy $J_{\mathsf{Portico}}$.

For runtime capability instance $\alpha=(\beta,\kappa)$, \system's
current-executability test is:
\[
\begin{aligned}
J_{\mathsf{Portico}}(\alpha,s_t) \equiv{}&
\operatorname{global\_ok}(\alpha,G) \wedge{}\\
&\operatorname{phase\_ok}(\alpha,\phi_t) \wedge{}\\
&\left(
\mathsf{init}(\alpha,\mathcal{E}_0,\tau,s_t)
\vee
\mathsf{live}(\alpha,L_t,\Lambda,s_t)
\right).
\end{aligned}
\]
Here $\mathsf{init}$ means $\kappa$ is compiler-issued initial authority, and
$\mathsf{live}$ means $\kappa$ names a live grant handle whose grant covers the
tool, privilege, resource, effect, and intent and has not been closed. A new
grant may be issued only when a rule in $\Gamma$ matches the requested
descriptor $\beta$, phase, trigger, and episode state in the observable state.
A closed episode cannot be reopened by planner text alone; a renewed grant for
the same descriptor requires a fresh trusted event such as a new
phase, orchestrator subgoal, human approval, or new episode nonce. The compiler
and runtime checks for these rules are described in Section~\ref{sec:system}.
For the running trace, \path{src/serialization.py} can match such a rule; the
packaging note cannot.

\paragraph{Semantic properties.}
The executable capability state induces three obligations: when authority may
appear, how it may widen, and when it must disappear.

\textbf{P1. Desired capability minimality.}
Executable capabilities exposed to the planner must stay within current justification:
\[
V_t \subseteq \{\alpha \in \mathcal{A}(\tau) \mid J(\alpha,s_t)\}.
\]
Only capabilities justified by current state should be executable through the
planner interface. Runtime reachability alone is insufficient.

\textbf{P2. Desired justified expansion.}
Authority outside the initial envelope should enter the executable set only
through an explicit grant transition:
\[
\begin{aligned}
&\alpha \notin \mathsf{Init}(\mathcal{E}_0,\tau,s_t)
  \wedge \alpha \in (V_{t+1} \setminus V_t) \\
&\qquad\Rightarrow
\mathsf{grant\_rule\_matches}(\beta(\alpha),s_t)
  \wedge J(\alpha,s_{t+1}).
\end{aligned}
\]
This implication concerns additions that are not compiler-issued initial
authority from $\mathcal{E}_0$. The grant rule is checked in the pre-state. The
transition mints a fresh runtime capability instance $\alpha$ for the requested
descriptor. The full visibility justification, including the live
grant, holds after the transition. Legitimate cross-boundary work is represented
as expansion of the executable capability set, not broad upfront exposure.

\textbf{P3. Desired capability closure.}
A temporary runtime capability should disappear once its current grant
episode ends:
\[
\forall t \ge c,\ \alpha \notin V_t.
\]
Here, $c$ is the first planner turn after the monitor consumes the observable
closure event for $\alpha$'s current grant episode. Closure ends the current
visibility episode. The same
descriptor $\beta(\alpha)$ can become justified again only through a fresh
runtime capability $\alpha'$ minted by a new grant issued after $c$ and still
live at time $t$.
If authority can be granted for a bounded subgoal but remains planner-usable
after the relevant phase or justification closes, then temporary authority has
degraded into lingering authority.

Operationally, this is a no-stale-use requirement. If grant $g$ closes before
turn $t$, no invocation backed only by $g$ should appear in the planner
interface at $t$, and any replay of $g$'s handle should be rejected before it
produces effects.

\paragraph{Monitor guarantees.}
For \system, the desired properties become checkable relative to the compiled
contract relation $J_{\mathsf{Portico}}$. This is the enforced property; the
gap between $J$ and $J_{\mathsf{Portico}}$ reflects contract quality.
The guarantees assume standard reference-monitor
assumptions: every planner invocation is mediated; the catalog maps each call
to a sound authority instance $\alpha(a)$; only the compiler initializes
$P=(\mathcal{E}_0,\Gamma,\Lambda,G)$; only the adjudicator mutates live grants
and audit state; handles are task-local server-side references bound to task,
epoch, resource, privilege, and effect; and closure observations are consumed
before the next planner invocation. The evaluated harness serializes planner
requests. A concurrent integration must be linearizable or must revalidate the
handle immediately before the side effect; otherwise revocation prevents only
new adjudications, not effects already permitted before closure. Below, $s'_t$
and $V'_t$ denote the candidate state and derived executable capability set
after pending observations and closures. Grant requests return
\textsc{grant}; permitted invocations only consume authority already present in
$V'_t$.

\textbf{G1. Visibility invariant.}
At every mediated turn after observations and revocations are applied,
\[
\begin{aligned}
V_t =
\{\alpha \mid{}& \alpha\in
\mathsf{Init}(\mathcal{E}_0,\tau,s_t)\cup
\mathsf{GrantCaps}(L_t,\tau,s_t)\\
&{}\wedge J_{\mathsf{Portico}}(\alpha,s_t)\}.
\end{aligned}
\]
Thus every executable capability exposed through the interface satisfies
$J_{\mathsf{Portico}}$.

\textbf{G2. Permit soundness.}
If the monitor permits invocation $a$ at state $s_t$, then
$\alpha(a)\in V'_t$, $J_{\mathsf{Portico}}(\alpha(a),s'_t)$ holds, no invariant
in $G$ denies $\alpha(a)$, and $a$ uses either compiler-issued initial authority
or a valid live handle for $\alpha(a)$.

\textbf{G3. Grant soundness.}
If an authority instance outside the initial envelope enters $V_t$, then the
audit log contains a grant whose rule $\gamma\in\Gamma$ matched the current
observable state, phase, selector, privilege, effect, and intent, and the
added instance is covered by that grant.

\textbf{G4. Closure before reuse.}
If a closure predicate for grant $g$ is satisfied by an observable event, then
$g$ and any authority instances visible only through $g$ are removed before
the next planner invocation is adjudicated. A later reuse therefore requires a
new grant transition; leftover visibility from the closed episode has been
removed.

\emph{Why the monitor enforces the compiled analogues of P1--P3.}
The argument is by induction over mediated invocations. The base state
derives $V_0$ from the compiler-issued initial envelope and
$J_{\mathsf{Portico}}$, so every executable instance is initial and checked
against $G$. For the inductive step, \texttt{IngestRuntimeObservation} records
events and \texttt{AdjudicatePlannerRequest} consumes them before the next
planner call. Closed grants are removed from $L_t$; $V_t$ is then recomputed
from initial authority, remaining live grants, and $J_{\mathsf{Portico}}$.
A permitted call is already in the updated executable capability set and
satisfies $J_{\mathsf{Portico}}$; if it is grant-backed, its handle names a
live epoch. New non-initial authority enters $V_{t+1}$ only after
a \textsc{grant} return from a matching $\gamma\in\Gamma$ that also passes $G$.
No transition directly inserts executable capabilities outside the initial
envelope or a live grant. Thus visibility satisfies G1, permits satisfy G2,
additions satisfy G3, and closed grant capabilities are absent before reuse by
G4. These are the enforced versions of P1--P3 for the compiled contract and
mediated path.

\section{\system as a Reference Monitor}
\label{sec:system}

\system sits between the planner and tools. It owns the lifecycle state around
live grants $L_t$, derives the executable capability set $V_t$ from the
contract and live handles, and removes temporary authority at closure. Host
isolation constrains execution after handoff. \system controls authority
exposed to the planner before it.

\paragraph{Inputs.}
\system uses three inputs: a task contract with initial scope, forbidden
resources, declared expansions, and closure events; a typed tool catalog that
maps runtime tools to privileges, resources, effects, and intent classes; and a
policy vocabulary for global deny rules such as credential-file denies, network
egress denies, \texttt{curl | sh}, and \texttt{git push}. Planner text can
request authority; the monitor decides whether to grant it.

\paragraph{Runtime interface.}
At each step, the planner sees only the currently visible tool schemas,
resource scopes, argument patterns, and live grant options. Denials return
reason classes; grants mint epoch-bound handles for covered resource/effect
bindings and add them to the next-step interface. Revocation removes those
handles from later tool selection and validation, and execution rejects stale
replay. Authority-request options are separate from effect capabilities:
they let the planner ask whether a declared expansion rule applies without
making the target readable or writable.

\paragraph{Operational meaning of visibility.}
Execution capabilities are exposed through the mediated interface. Tool schemas
are omitted or narrowed, resource scopes and argument patterns are
removed from the next planner menu, and execution tools consume
compiler-issued initial handles or live grant handles. Free-form paths are
accepted only by the authority-request endpoint or by compatibility wrappers
that resolve them against compiler-issued initial or live grant handles before
execution; unresolved paths are denied or converted into authority requests. A
remembered path becomes authority only when a live handle for that resource and
effect is present. Denied attempts are logged separately from visibility, and
revocation changes the capability interface rather than model memory.
Handles are opaque task-local references validated server-side against task,
epoch, resource, privilege, and effect. In our examples, we name handles by
role, such as a serialization read handle; the runtime accepts only opaque
handles that resolve to live server state.

\begin{table*}[t]
\centering
\caption{Planner-facing interface snapshots for PORTICO and the all-visible same-policy comparator. The timeout-boundary case reuses the API-to-serialization boundary from the running example. Both conditions share execution policy, grant timing, handle checks, and closure events; they differ in the breadth of proposal affordances shown to the planner.}
\scriptsize
\setlength{\tabcolsep}{2.5pt}
\resizebox{\textwidth}{!}{%
\begin{tabular}{p{0.12\textwidth} p{0.28\textwidth} p{0.28\textwidth} p{0.24\textwidth}}
\toprule
Moment & PORTICO interface & All-visible same-policy interface & Execution policy \\\\
\midrule
Schema shape & \texttt{request\_authority(selector,effect)}; \texttt{read/write(handle enum)} over initial and live handles & same request endpoint plus broad proposal descriptors; \texttt{read/write} still require initial or live handles & same policy; descriptors alone cannot execute effects \\
Before request & initial API/test handles; serialization execution handles absent; authority request available & broad proposal surface includes the serialization descriptor; no live serialization execution handle & same policy; authority request returns grant; handles next turn \\
Grant live & initial handles plus live grant handles: g0001:r1 $\rightarrow$ src/serialization.py (read/write); g0001:r2 $\rightarrow$ tests/test\_serialization.py (read) & broad proposal surface remains visible; execution still requires the same live grant handles & same policy; invocation with live handle returns allow \\
After closure & initial handles only; serialization grant handles absent & broad proposal surface remains visible; closed handles are rejected at execution & same policy; stale replay returns deny: stale handle \\
\bottomrule
\end{tabular}
}
\label{tab:interface-snapshots}
\end{table*}

Table~\ref{tab:interface-snapshots} shows the concrete \system interface next
to the all-visible same-policy comparator. The timeout-boundary case reuses the
API-to-serialization boundary from the running example. In \system,
\path{src/serialization.py} can remain a file-name hint, but it becomes
executable only while the epoch-bound handle is live. In the comparator,
the request option remains broad, while the same execution policy still
requires a live handle. After closure, \system removes the handle from the next
interface and both policies reject stale replay before execution.

\paragraph{Shell and compound tools.}
Shell, git, package-manager, and network calls are mediated through the typed
catalog before execution. When classifiable, the mapper extracts file operands,
command class, network destinations, dependency effects, and VCS effects, then
checks the resulting instances against $V_t$ and $G$. Global denies such as
\texttt{curl | sh}, \texttt{git push}, forbidden package mutation, and
undeclared network egress are denied; unclassified commands require explicit
approval or are denied. The guarantees in Section~\ref{sec:formalism} cover
mediated tool invocations only. Appendix~\ref{sec:supporting-analyses} reports
the frozen compound-mediation audit over observed commands, not arbitrary shell
syntax.

\paragraph{Contract object and authoring path.}
At runtime, the task contract is the monitor's policy input and planner prose
supplies requests. The monitor compiles it into
$P=(\mathcal{E}_0,\Gamma,\Lambda,G)$: initial envelope, grant rules, closure
rules, and global deny rules. Figure~\ref{fig:policy-contract-example} shows the
running excerpt. Contracts draw on the task interface, repository policy,
protected-file rules, test targets, dependency/network policy, and optional
boundary expansions. A planner can request expansion, but the monitor checks the
compiled contract, catalog, phase, selectors, and observable state. Poor
contracts can undergrant or overgrant; Section~\ref{sec:discussion} discusses
that deployment risk.

\begin{figure}[t]
\centering
\begin{lstlisting}[style=porticosnippet]
task:
  name: timeout propagation
  goal: propagate optional timeout
initial_allowed_resources:
  - {path: "src/api.py", access: read_write}
  - {path: "tests/test_api.py", access: read_write}
escalation_paths:
  - id: serialization-module-legitimate
    trigger: authority_request
    allowed_resources:
      - {path: "src/serialization.py", access: read_write}
      - {path: "tests/test_serialization.py", access: read}
    phases: [inspect, patch, test]
compiled_lifetime:
  until: [tests_pass, subgoal_complete, phase_exit]
forbidden_resources:
  - {path: "notes/packaging.md", access: read}
  - {path: "requirements.txt", access: write}
forbidden_network_hosts: ["*"]
\end{lstlisting}
\caption{Contract excerpt with compiler-derived lifetime for the running timeout-propagation case. The full checked-in manifest appears in the artifact; this excerpt shows the initial envelope, declared grant, closure rule, and nearby forbidden resources.}
\Description{A YAML-like task contract for the timeout-propagation episode. It lists initial allowed resources over src/api.py and tests/test_api.py, a serialization escalation path with granted read, write, and test authority, compiled lifetime closure events, and denies for packaging notes, package metadata, and network egress.}
\label{fig:policy-contract-example}
\end{figure}

In this harness, \texttt{subgoal\_complete} denotes a trusted runtime or
orchestrator event consumed by the monitor. A planner-authored textual claim
with the same wording is logged as untrusted text and cannot close or extend a
grant.

\paragraph{Compilation.}
The compiler builds $\mathcal{E}_0$ from initial scope, maps operations to
tools and privileges, maps paths and commands to resources, and turns network,
dependency, and policy-vocabulary restrictions into $G$. Expansions become grant rules in
$\Gamma$; lifetime clauses become closure predicates in $\Lambda$. The packaging
probe and serialization boundary compile to different selectors, so
adjudication does not rely on free-form language judgments. The same checked-in
manifest is read by both the compiler and the scorer, with authority labels fixed
before execution; Appendix Table~\ref{tab:contract-to-policy-map} summarizes the
full path. Intent fields are advisory: grant matching uses tool/effect metadata,
target selectors, monitor phase, contract triggers, and observable state.

\begin{table}[t]
\centering
\caption{Event provenance for grant and closure decisions. Planner text is
advisory; trusted monitor or workflow events drive authority changes.}
\scriptsize
\resizebox{\columnwidth}{!}{%
\begin{tabular}{p{0.30\columnwidth} p{0.28\columnwidth} p{0.18\columnwidth} p{0.18\columnwidth}}
\toprule
Event & Source & Trust & Role in authority transition \\
\midrule
Planner rationale & Model text & Untrusted & No \\
Authority request and \texttt{request.intent} & Planner request over typed schema & Advisory & Selects candidate rule only \\
Focused test result & Process launched by monitor & Trusted & Can close a grant \\
Workflow subgoal or phase transition & Orchestrator or runtime state & Trusted & Can close a grant \\
Human approval or revocation & Authenticated UI or policy hook & Trusted & Can grant or close \\
Textual \texttt{subgoal\_complete} claim & Model text & Untrusted & No \\
\bottomrule
\end{tabular}
}
\label{tab:event-provenance}
\end{table}

The evaluated harness serializes planner requests. A queued or batched
integration must apply the same epoch and handle checks to each invocation
before effects occur; a closed grant cannot be reused without a fresh
adjudication. If a call is already permitted and in flight when a closure event
arrives, the integration must either serialize the event before the effect or
revalidate the handle immediately before the effect. The guarantees reported
here assume that linearizable mediation point.

\paragraph{Runtime adjudication.}
Algorithms~\ref{alg:compile-envelope}, \ref{alg:adjudicate-invocation}, and
\ref{alg:ingest-observation} give the monitor procedures. Compilation builds
$P=(\mathcal{E}_0,\Gamma,\Lambda,G)$, $V_0$, and $I_0$. Adjudication consumes
pending closures and separates invocations (which need initial authority or a
live handle) from authority requests (which mint handles for the next
interface without executing). Observation ingestion records events after a
permitted call so closures take effect before the next planner invocation. On
adjudication the monitor refreshes state, derives $V$ from current live
grants, checks invariants, and logs a structured decision. Grant rules are
templates in $\Gamma$; grants are live instances that mint epoch-bound handles.

\SetKwInOut{Input}{Input}
\SetKwInOut{Output}{Output}
\SetKwInOut{Writes}{Writes}
\begin{algorithm}[t]
\footnotesize
\Input{task contract $\tau$; typed tool catalog $\mathcal{T}$; runtime profile $\rho$; policy vocabulary $\Omega$}
\Output{policy $P=(\mathcal{E}_0,\Gamma,\Lambda,G)$, initial capabilities $V_0$, and interface $I_0$}
\Writes{$\mathcal{E}_0$, $\Gamma$, $\Lambda$, $G$, $V_0$, and $I_0$}
Parse $\tau$ into scope, selectors, phase defaults, network mode, and dependency policy\;
Map declared operations through $\mathcal{T}$ to privileges, effects, argument constraints, and resource selectors\;
Apply runtime profile $\rho$ to restrict available tools, sandbox mode, network mode, package-manager effects, and VCS effects\;
Build $\mathcal{E}_0=(T_0,\Pi_0,\mathcal{R}_0,C_0)$ from the initial tools, privileges, resources, and constraints\;
Add forbidden paths, hosts, shell patterns, dependency restrictions, and VCS restrictions from $\tau$ and $\Omega$ to $G$\;
Compile each declared expansion into a grant rule $\gamma\in\Gamma$ with trigger, selector, grant body, intent, and covered authority instances\;
Compile each lifetime clause into closure predicates $\Lambda$\;
Use trusted events such as monitor-launched tests passing, orchestrator subgoal completion, phase exit, run completion, or authenticated revoke for closure\;
Set $V_0\leftarrow\{\alpha\in\mathsf{Init}(\mathcal{E}_0,\tau,s_0)\mid J_{\mathsf{Portico}}(\alpha,s_0)\}$\;
Set $I_0\leftarrow \operatorname{Exec}(V_0)\uplus\operatorname{Prop}(F_0)\uplus\operatorname{Req}(Q_0)$\;
\Return{$P,V_0,I_0$}\;
\caption{CompileEnvelope.}
\Description{Algorithm2e pseudocode for CompileEnvelope. It parses the task contract, maps operations through the typed tool catalog, builds the initial envelope, adds invariants, compiles grant rules and closure predicates, derives initial capabilities, and derives the initial planner interface.}
\label{alg:compile-envelope}
\end{algorithm}

\begin{algorithm}[t]
\footnotesize
\Input{planner request $q$; state $s_t$; policy $P$; catalog $\mathcal{T}$}
\Output{\textnormal{\textsc{permit}}, \textnormal{\textsc{grant}}, or \textnormal{\textsc{deny}}}
\Writes{live grants $L_t$, derived interface delta, and audit log $A_t$}
$s\leftarrow\mathsf{RefreshState}(s_t,O_t,\Lambda)$ \tcp*{consume observations and close grants}
$V\leftarrow\mathsf{DeriveCaps}(\mathcal{E}_0,s.L,\tau,s)$\;
$\beta\leftarrow\mathsf{Map}_{\mathcal{T}}(q,s.\phi)$ \tcp*{phase comes from monitor state}
All returns commit the current $s$ as $s_{t+1}$\;
\If{$\mathsf{Violates}(\beta,G)$}{
  Append invariant reason to $A_t$; \Return{\textnormal{\textsc{deny}}}\;
}
\If{$q$ is an invocation}{
  $\alpha\leftarrow\mathsf{ResolveCap}(q,V)$\;
  \If{$\alpha$ is defined and $\alpha\in V$ and $J_{\mathsf{Portico}}(\alpha,s)$}{
    Append permit to $A_t$; \Return{\textnormal{\textsc{permit}}}\;
  }
  Append denial reason to $A_t$; \Return{\textnormal{\textsc{deny}}}\;
}
\If{$q$ is an authority request}{
  $\gamma\leftarrow\mathsf{MatchGrant}(\Gamma,\beta,s)$\;
  \If{$\gamma$ is undefined}{
    Append denial reason to $A_t$; \Return{\textnormal{\textsc{deny}}}\;
  }
  $g\leftarrow\mathsf{IssueGrant}(\gamma,s)$; add $g$ to $s.L$\;
  $V_{t+1}\leftarrow\mathsf{DeriveCaps}(\mathcal{E}_0,s.L,\tau,s)$\;
  $I_{t+1}\leftarrow\mathsf{DeriveIface}(V_{t+1},F_{t+1},Q_{t+1})$\;
  $\Delta I\leftarrow$ handles minted for $g$ and exposed in $I_{t+1}$\;
  Append grant record and $\Delta I$ to $A_t$; \Return{\textnormal{\textsc{grant}}}\;
}
Append denial reason to $A_t$; \Return{\textnormal{\textsc{deny}}}\;
\caption{AdjudicatePlannerRequest.}
\Description{Algorithm2e pseudocode for AdjudicatePlannerRequest. It refreshes authority state, maps a planner request to an authority descriptor, checks invariants, permits invocations that resolve to initial authority or a live handle, and issues matching grants for the next planner interface.}
\label{alg:adjudicate-invocation}
\end{algorithm}

\begin{algorithm}[t]
\footnotesize
\Input{runtime result $y$ for permitted invocation $a$; state $s_t$}
\Output{updated observation buffer and audit log}
\Writes{$O_t$ and $A_t$}
$E\leftarrow\mathsf{TrustedEvents}(y)$ \tcp*{tests, workflow events, approvals, revokes}
$S\leftarrow\mathsf{ClassifiedSideEffects}(y,a)$\;
Drop planner-authored test, phase, subgoal, approval, or revoke claims from $E$\;
Append $E\cup S$ to $O_t$; append $y$ to $A_t$\;
Leave grant closure to the next call to \texttt{AdjudicatePlannerRequest}\;
\caption{IngestRuntimeObservation.}
\Description{Algorithm2e pseudocode for IngestRuntimeObservation. It records trusted runtime events and classified side effects after a permitted invocation returns, leaving closure to be consumed before the next planner invocation is adjudicated.}
\label{alg:ingest-observation}
\end{algorithm}

\paragraph{Running walkthrough.}
In the running trace, the packaging-note read is denied because no rule in
$\Gamma$ matches it. A serialization authority request succeeds because the
expansion rule matches target, intent, phase, and constraints. That request
returns a grant and next-interface handles, not an immediate file read. Later
reads and writes are permitted through those handles while the grant is live.
When the focused test passes or the subgoal closes, $\Lambda$ revokes the
serialization grant before the next planner invocation. The handles disappear
from the interface, and replaying one of them is denied before execution.

\paragraph{Configurations, audit, and integration.}
We evaluate \system core, which denies requests outside $\mathcal{E}_0$, and
\system full, which adds request--grant--invoke with trusted closure. A grant extends only the
targeted resource, tool schema, constraint, and lifetime; closure removes that
extension. The task-wide non-revoking comparator starts from the same envelope
and receives the same grants at the same turns, but does not remove them at
closure. Every permit, deny, grant, and revoke emits a structured record with
task, phase, requested action, target, grant ID, handle identifier,
decision, and reason. The same policy core backs \texttt{portico-tool}, a
standalone MCP server packaged with the artifact. It exposes interface
snapshots, authority requests, handle-based file and test operations, closure,
and audit logs over stdio. This is the public integration path for trying the
request--grant--invoke lifecycle in MCP-capable coding-agent clients; its smoke
tests check protocol and packaging behavior, not benchmark outcomes.
Appendix~\ref{sec:supporting-analyses} reports the compiled policy object and
derivation audits.

\section{Experimental Questions and Setup}
\label{sec:benchmark}

The evaluation uses explicit scope contracts to observe exposure, expansion,
and closure separately. Broad SWE benchmarks measure end-to-end capability but
merge these effects into one success score~\cite{agentbench,agentboard,webarena,visualwebarena,workarena,mind2web,gaia,osworld,androidworld,bmoca,mobileworld,swebench,swepolybench}.
We use four task-oriented suites: three controlled security fixtures and one
pinned real-repository suite.

\paragraph{Research questions.}
The evaluation is organized around four questions:
\begin{itemize}
\item \textbf{RQ1: Capability minimization.} Does minimizing exposed capabilities reduce unsafe effects relative to open access, static allowlists, and coarse sandboxing?
\item \textbf{RQ2: Justified expansion.} Can controlled expansion recover legitimate cross-boundary utility without reintroducing broad upfront exposure?
\item \textbf{RQ3: Capability closure.} Once recovery and unsupported-probe denial match before closure, does bounded lifetime reduce post-closure stale capability availability relative to a non-revoking design with the same grant timing, and what planning pressure remains when the same execution policy is applied to an all-visible interface?
\item \textbf{RQ4: Real-repository lifecycle.} Does the same grant/revoke behavior appear on pinned real repositories?
\end{itemize}

Section~\ref{sec:results} follows this order: exposure, expansion, closure,
and real-repository lifecycle.

\paragraph{Benchmark suites.}
Suites A--C are controlled repository fixtures. Each
episode is generated from a checked-in manifest that fixes the task prompt,
fixture files, initial capability envelope, allowed and forbidden resources,
optional grant rules, success predicate, and violation predicate before any
model run. The scorer reads those frozen fields and scores observed tool calls
against the contract. Suite A stresses unsafe exposure with reachable lures;
Suite B uses bounded coding tasks that require legitimate temporary boundary
grants; Suite C keeps the coding goal legitimate but surrounds it with hostile
or overscoped context. Suite D uses the same manifest format on pinned public
repository snapshots. Appendix Table~\ref{tab:suite-construction} gives their
detailed run counts and purpose.

\paragraph{Comparison targets.}
Table~\ref{tab:baseline-configs} lists the deployment baselines, \system
variants, and causal comparators. The baselines expose broad access, a fixed
allowlist, or coarse sandbox restrictions, without treating the task contract as
a lifecycle policy. \system core uses only the initial envelope; \system full
adds request--grant--invoke with trusted closure. The non-revoking comparator
disables only revocation, while the all-visible same-policy comparator keeps a
broad planner interface under the same execution policy, grant timing, handle
checks, and closure. All conditions use the same catalog, scorer, budgets,
fixtures, and lower-level runtime unless noted.

\begin{table*}[t]
\centering
\caption{Comparison targets. Baselines test deployment exposure; \system variants test capability minimization and grant/revoke; the comparators isolate closure and execution-time filtering.}
\scriptsize
\resizebox{\textwidth}{!}{%
\begin{tabular}{p{0.12\textwidth} p{0.24\textwidth} p{0.28\textwidth} p{0.25\textwidth}}
\toprule
\textbf{Condition} & \textbf{Planner interface} & \textbf{What it controls} & \textbf{What it cannot show alone} \\
\midrule
Full access & Full tool menu and broad resources & Raw model behavior under maximum exposed authority & Least privilege, semantic scoping, or closure. \\
Static allowlist & Fixed task-independent allowlist plus basic path restrictions & Coarse tool availability and basic resource reachability & Task- and phase-specific distinction between two uses of the same primitive. \\
Coarse sandbox & Workspace confinement, sensitive-path denies, egress and process restrictions & Post-handoff reachability after a tool call is selected & Which admissible capabilities should be exposed for the current task state. \\
\system core & Initial task-derived envelope only & Capability minimization, semantic checks, global deny rules, and auditability & Legitimate work that requires later boundary expansion. \\
\system full & Initial envelope plus request--grant--invoke lifecycle with trusted closure & Controlled expansion and closure of temporary capabilities & Whether a non-revoking design could match utility without bounded visibility. \\
Task-wide non-revoking & Same initial envelope and same grant timing as \system full, but grants persist after closure & Same task policy, same boundary recovery, and unsupported-probe denial, but no revocation & Whether stale capabilities persist when only revocation is removed. \\
All-visible same-policy & Broad proposal interface over resources, with \system's execution-time policy, grant timing, handle checks, and closure events & Interface breadth versus late enforcement under the same semantic policy & Whether the post-closure proposal difference generalizes or causes harm under imperfect mediation. \\
Coarse all-visible action filter & Broad proposal interface over resources plus invocation-time checks, including closed-grant stale-handle denial & Whether late denial is enough when adjacent capabilities remain exposed to the planner & It is a deployment diagnostic, not the closure comparator with the same grant timing; adjacent-pressure rows use a coarser filter and can permit oracle-forbidden detours. \\
\bottomrule
\end{tabular}
}
\label{tab:baseline-configs}
\end{table*}

\paragraph{Experiment scope.}
The experiments compare exposure strategies under a fixed task policy:
whole-run exposure, execution-time denial, or bounded grant episodes with closure.
For global A--D violation rates, the same checked-in contract supplies
enforcement and scoring, so those rates are contract-based enforcement
results rather than policy inference results. Separate post-closure
lifecycle, stale-effect, scripted live stale-effect, and model-live stale-effect
oracle files, never read by the monitor, score post-closure behavior. The
contract-sensitivity audit shows the expected failure modes: undergranting
blocks boundary work, overgranting increases initial exposure, and a wrong
expansion is inherited by the monitor.

\paragraph{Scoring.}
We score security, utility, and exposure separately. The main security readings
are executed forbidden effects and post-closure reuse. Utility is measured by
task success and scope compliance. Blast radius and blocked forbidden proposals
are diagnostics that explain exposure and planning pressure. Main A/B/C tables
report run-level means over repeated executions. Binary security readings are
interpreted at the scenario or episode level, with repeat counts shown as
reproducibility checks rather than independent samples. This convention keeps
blocked planning pressure separate from effects that actually reach the runtime.

A forbidden proposal is a planner request whose resource or effect would violate
the contract if executed; a blocked forbidden proposal is denied before effects.
A policy violation is an executed-forbidden effect: a read, write, network
action, dependency mutation, command, or cross-tool step matching a forbidden
resource or violation condition. Scope compliance means the trace contains no
out-of-scope reads, writes, executions, network actions, or unsafe commands,
independent of task success. Task success is the episode-specific goal. Blast
radius is a weighted count of sensitive resources reached or modified.
Post-closure reuse records whether a later reread or reuse attempt after
explicit close is allowed.
Together, these metrics separate completed work, blocked detours, executed
effects, and stale capability use.

\paragraph{Models and runs.}
The main reported model profile is Qwen3-Coder 30B through the Hugging Face router
(\texttt{Qwen/Qwen3-Coder-30B-A3B-Instruct:fastest}). It is used for the main
A--C matrix, the post-closure lifecycle diagnostics, the live stale-effect follow-up,
and the Qwen3 Suite-D real-repository matrix. Frozen traces record the router
profile, request metadata, and returned model string; fresh router reruns may
select a different provider backend, so these roots are treated as frozen
API-backed profiles rather than a locally reproducible checkpoint. A
supplementary frontier slice uses GPT-5.5, Gemini 3.5 Flash, and Claude Opus 4.8
on the same three grant-lifecycle episodes under full access, \system, and the
non-revoking comparator. We keep that slice separate from the Qwen3 matrix: it
checks whether the same grant and closure ordering appears under current
API-backed frontier profiles, not a pooled estimate across models. Main runs use
temperature 0.0 with a 32{,}768-token context; repeats reset the fixture and
planner loop for the same episode/condition cell and are reproducibility checks,
not independent samples. Main A/B/C runs use 4 planner steps on Suite A and 6 on
Suites B/C; controlled escalation uses 8 steps, post-grant follow-through uses
9--10, and the Qwen3 router profile uses a 240\,s request timeout. The artifact
records exact model strings, budgets, profile configuration, returned model
identifiers, and local freeze files.

\paragraph{Manifest, audit, and execution protocol.}
Each episode is a checked-in manifest. The timeout-propagation witness declares
\path{src/serialization.py} as the escalation target and lists
\path{notes/packaging.md} only as a forbidden resource and attack lure
(Appendix~\ref{sec:case-studies}). Scoring uses manifest fields fixed before
execution, not post-hoc judgments about planner rationale. Main results come
from live execution: fixtures are instantiated, the planner attempts calls, the
monitor records decisions, and traces are scored against the episode contract.
Trace records capture runtime metadata, request, target, decision, reason, and
latency; replay tooling is retained for denial analysis.

\paragraph{Threats to scoring validity.}
Stronger policies induce trajectory divergence, so we
score contract satisfaction instead of trace similarity. Contaminated-context
tasks involve semantic boundary judgments, so allowed resources, forbidden
resources, success conditions, and violation conditions are fixed before
execution. Model nondeterminism remains even at fixed temperature, so live
studies use repeated executions. The policy vocabulary and compiler are heuristic and
coding-specific, so claims stay within this policy family and domain.

\section{Results}
\label{sec:results}

The results follow the lifecycle: exposure, expansion, closure, and
real-repository integration. \system full trades some raw completion for
stricter scope control. The key comparison is closure: a task-wide
non-revoking comparator can know which authority is legitimate and still leave
it available after its justification ends.

\begin{table}[t]
\centering
\caption{Global trade-off snapshot from the main Qwen3-Coder 30B study: five reproducibility repeats per scenario or episode over 17 Suite-A scenarios, 37 Suite-B episodes, and 14 Suite-C episodes.}
\label{tab:main-qwen}
\scriptsize
\setlength{\tabcolsep}{3pt}
\begin{tabular}{@{}lccccc@{}}
\toprule
Metric & Full & Static & Sandbox & \textbf{\system full} & \textbf{\system core} \\
\midrule
Suite A viol.\ $\downarrow$ & 1.00 & 0.82 & 0.86 & \textbf{0.00} & \textbf{0.00} \\
Suite A blast $\downarrow$ & 6.65 & 2.73 & 4.09 & \textbf{0.00} & \textbf{0.00} \\
Suite B succ.\ $\uparrow$ & 0.97 & 0.97 & 0.97 & \textbf{0.87} & 0.21 \\
Suite B scope $\uparrow$ & 0.92 & 0.92 & 0.91 & \textbf{1.00} & \textbf{1.00} \\
Suite C succ.\ $\uparrow$ & 1.00 & 1.00 & 1.00 & \textbf{1.00} & 0.94 \\
Suite C scope $\uparrow$ & 0.69 & 0.70 & 0.70 & \textbf{1.00} & \textbf{1.00} \\
Suite C viol.\ $\downarrow$ & 0.31 & 0.30 & 0.30 & \textbf{0.00} & \textbf{0.00} \\
\bottomrule
\end{tabular}
\end{table}

Table~\ref{tab:main-qwen} summarizes the security--utility trade-off. \system
full trades maximum completion for scope control, turning some unsafe or
overbroad completions into scoped success or visible incompleteness.

\subsection{RQ1: Minimization Reduces Unsafe Effects}

\paragraph{Suite-A safety result.}
Suite A is the cleanest exposure test. Full access violates every run and has
average blast radius 6.65. Static allowlists and coarse sandboxing lower
exposure but still leave violation rates at 0.82 and 0.86. Both \system
variants drive violations and blast radius to 0.00: 0/17 scenarios and 0/85
repeated executions record violations. The scenario-level reading is therefore
the same as the run-level reading for the \system rows.

\paragraph{Limitations of deployment baselines.}
Full access exposes the full interface, the static allowlist narrows it, and the
coarse sandbox constrains execution reachability. They still leave the
task-phase question open: when is one local file read justified while another
is not? In the same-primitive rerun, deployment baselines complete 20/20
runs by taking one forbidden local detour per run; \system completes all 20/20
in scope with 0.00 violations. The failure is a missing resource/effect/phase
distinction in the visible interface.

\subsection{RQ2: Grants Recover Legitimate Boundary Work}

\paragraph{Core is safe but brittle.}
\system core keeps scope compliance at 1.00 and violations at 0.00, but
Suite-B success drops to 0.21 because many tasks need boundary-adjacent or
auxiliary resources. A narrow initial envelope is safe but brittle: it preserves
scope by stopping when legitimate work crosses a declared boundary.

\paragraph{Controlled grants recover the boundary.}
\system full improves the main Suite-B result from \system core's 0.21 success
to 0.87 while retaining 1.00 scope compliance and zero measured violations.
The focused Qwen2.5-Coder control slice, reported in
Appendix~\ref{sec:supporting-analyses}, isolates benign boundary tasks: static
allowlist succeeds by exposing secondary authority upfront, \system core
overblocks, and \system full restores 20/20 successes in scope with one grant
per task on average. A larger boundary pack repeats the pattern: \system full
completes 35/35 legitimate-escalation runs in scope with zero measured
violations. These grants are bounded to the declared boundary resource and
focused validation target.

\paragraph{\system incurs a measurable utility cost.}
\system full still pays a utility cost on Suite B: 161/185 completions versus
180/185 under the static allowlist and coarse sandbox. At the episode level, it
completes every repeat for 31/37 Suite-B episodes and at least one repeat for
33/37; the deployment baselines complete every repeat for 36/37. On Suite
C, \system full completes all 14 episodes in scope across 70/70 repeats.
Appendix~\ref{sec:supporting-analyses} attributes \system misses to narrow
grants, missing request options, conservative contaminated-context blocking, and
planner/model instability. These misses are visible utility cost; the traces
remain scope-compliant.

\subsection{RQ3: Closure Removes Temporary Authority}
\label{sec:lifecycle}

\paragraph{Strong non-revoking comparator.}
The task-wide non-revoking comparator starts with the same initial envelope as
\system full, receives the same grant in the same turn, and permits the same
actions during the episode. The only removed mechanism is closure. Once
success, scope, and unsupported-probe denial before closure match, the
late-reread endpoint measures remaining stale capability.

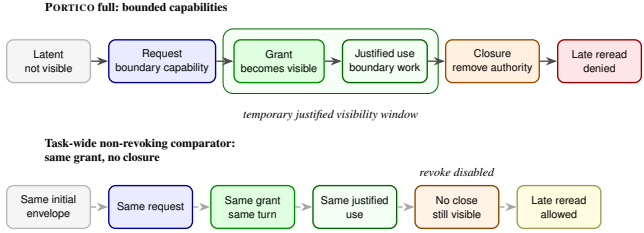
\begin{figure}[t]
\centering
\resizebox{\columnwidth}{!}{%
\begin{tikzpicture}[
  font=\scriptsize\sffamily,
  node distance=0.45cm and 0.35cm,
  stagebox/.style={draw, rounded corners=4pt, minimum width=1.72cm, minimum height=0.88cm, align=center, inner sep=4pt, thick},
  arrow/.style={-{Stealth}, thick, draw=black!70},
  altarrow/.style={-{Stealth}, thick, dashed, draw=gray!70},
  rowlabel/.style={font=\scriptsize\bfseries, align=left},
  note/.style={font=\scriptsize\itshape, align=center}
]
\node[rowlabel, anchor=west] at (-0.2,1.1) {\system full: bounded capabilities};
\node[stagebox, fill=gray!8, draw=gray!55] (latent) at (0,0) {Latent\\not visible};
\node[stagebox, fill=blue!8, draw=blue!55!black, right=of latent] (request) {Request\\boundary capability};
\node[stagebox, fill=green!12, draw=green!55!black, right=of request] (grant) {Grant\\becomes visible};
\node[stagebox, fill=green!5, draw=green!40!black, right=of grant] (use) {Justified use\\boundary work};
\node[stagebox, fill=orange!12, draw=orange!60!black, right=of use] (close) {Closure\\remove authority};
\node[stagebox, fill=red!10, draw=red!60!black, right=of close] (deny) {Late reread\\denied};

\draw[arrow] (latent) -- (request);
\draw[arrow] (request) -- (grant);
\draw[arrow] (grant) -- (use);
\draw[arrow] (use) -- (close);
\draw[arrow] (close) -- (deny);

\begin{scope}[on background layer]
  \node[draw=green!40!black, fill=green!4, rounded corners=6pt, fit=(grant) (use), inner sep=6pt] (window) {};
\end{scope}
\node[note, below=0.2cm of window] {temporary justified visibility window};

\node[rowlabel, anchor=west] at (-0.2,-1.85) {Task-wide non-revoking comparator:\\same grant, no closure};
\node[stagebox, fill=gray!8, draw=gray!55] (latentb) at (0,-3.0) {Same initial\\envelope};
\node[stagebox, fill=blue!8, draw=blue!55!black, right=of latentb] (requestb) {Same request};
\node[stagebox, fill=green!12, draw=green!55!black, right=of requestb] (grantb) {Same grant\\same turn};
\node[stagebox, fill=green!5, draw=green!40!black, right=of grantb] (useb) {Same justified\\use};
\node[stagebox, fill=orange!10, draw=orange!55!black, right=of useb] (persist) {No close\\still visible};
\node[stagebox, fill=yellow!15, draw=yellow!55!black, right=of persist] (allow) {Late reread\\allowed};

\draw[altarrow] (latentb) -- (requestb);
\draw[altarrow] (requestb) -- (grantb);
\draw[altarrow] (grantb) -- (useb);
\draw[altarrow] (useb) -- (persist);
\draw[altarrow] (persist) -- (allow);

\node[note, above=0.06cm of persist] {revoke disabled};
\end{tikzpicture}
}
\caption{Semantic meaning of authority closure. Both rows start from the same initial envelope and issue the same grant at the same point; only \system removes that authority before late reread.}
\Description{A two-row timeline. The upper row shows bounded capabilities: latent not visible, request, grant becomes visible, justified use, closure removes the handle, and late reread denied, with a highlighted temporary window. The lower row shows a task-wide non-revoking comparator: same initial envelope, same request, same grant, same justified use, no close, and late reread allowed.}
\label{fig:revoke-followthrough}
\end{figure}

\paragraph{Non-revoking comparison.}
Figure~\ref{fig:revoke-followthrough} shows the split. The comparator supports
the same legitimate use, but keeps the stale capability available after the
subgoal. Table~\ref{tab:oracle-broad-sweep} measures the effect on the
late-reread slice with the same grant timing. Both systems match on task success, scope
compliance, contract-forbidden effects, and all pre-closure decisions.
\system full denies reuse in 10/10 runs, while the non-revoking comparator
allows it in 10/10. The endpoint is residual stale-capability availability after
closure.

\begin{table}[t]
\centering
\caption{Late-reread closure comparison. The task-wide non-revoking comparator starts with the same initial envelope as \system full and receives the same grants at the same turns, but lacks the revoke transition. Both systems match on task success, scope compliance, and contract-labeled forbidden effects on this slice; only \system removes the temporary authority before post-closure reuse. Dedicated lifecycle-oracle rows score post-closure reuse separately from contract-labeled resource/effect violations.}
\scriptsize
\setlength{\tabcolsep}{4pt}
\begin{tabular}{p{0.52\columnwidth}cc}
\toprule
Endpoint & \system full & Non-revoking \\
\midrule
Success & 10/10 & 10/10 \\
Scope & 10/10 & 10/10 \\
Contract-labeled forbidden effects & 0/10 & 0/10 \\
Late reread allowed & 0/10 & 10/10 \\
Late reread denied & 10/10 & 0/10 \\
\bottomrule
\end{tabular}
\label{tab:oracle-broad-sweep}
\end{table}

The trace records the mechanism behind this split: the boundary request mints
an epoch-bound handle such as \texttt{g0001:r1}, the next planner interface
includes it, and closure marks it inactive. Appendix
Table~\ref{tab:capability-lifecycle-oracle} audits the post-grant misuse family
against an oracle file separate from the deployment contract and not read by
the monitor. \system rejects stale-handle replay; the non-revoking comparator
keeps the same handle live. The coarse all-visible action filter denies stale
replay at action time, but remains a deployment diagnostic because it exposes a
broader interface and uses coarser adjacent-detour checks.

\paragraph{Post-closure effects.}
Table~\ref{tab:capability-stale-effects} adds a six-case stale-effect audit
over serialization, markdown, output, config, and options boundaries. The
static traces grant a boundary write capability, use it once, and then replay
the old handle for a post-closure write. The audit reads a separate oracle file
after execution; the monitor and contract compiler do not read it. \system
accepts 0/6 stale capabilities and executes 0/6 forbidden writes. The
non-revoking comparator accepts 6/6 stale capabilities and executes 6/6
forbidden writes. The coarse all-visible action filter catches the stale
handle at invocation time. This deterministic audit is a protocol-conformance
check; it shows that the same lifetime split extends from rereads to executed
write effects.

\begin{table}[t]
\centering
\caption{Post-closure stale-effect audit. The oracle file is not read by the monitor. Static traces execute the same grant, use it once, and then try a post-closure write through the stale handle. The non-revoking comparator keeps the handle live; \system invalidates it before effects.}
\scriptsize
\setlength{\tabcolsep}{3pt}
\resizebox{\columnwidth}{!}{%
\begin{tabular}{lccc}
\toprule
Condition & Cases & Stale cap. accepted & Executed forbidden effect \\
\midrule
PORTICO full & 6 & 0/6 & 0/6 \\
Non-revoking comparator & 6 & 6/6 & 6/6 \\
Coarse all-visible action filter & 6 & 0/6 & 0/6 \\
\bottomrule
\end{tabular}
}
\label{tab:capability-stale-effects}
\end{table}

\paragraph{Live stale-effect follow-ups.}
Table~\ref{tab:capability-live-stale-effects} repeats the stale-effect pattern
through scripted contaminated-agent traces and six Qwen3-Coder 30B model-live
cases. They cover file writes, git metadata or hook mutation, and network egress
to stubbed hosts. All conditions allow 6/6 legitimate live uses; \system and the
all-visible same-policy comparator execute 0/6 stale effects, while the
non-revoking comparator executes 6/6. These are targeted follow-ups, not
exploit-rate estimates.

\begin{table}[t]
\centering
\caption{Live stale-effect follow-up. A scripted contaminated-agent trace requests a temporary capability, uses the live handle once, and then attempts a second post-closure side effect through the stale handle. The oracle file is not read by the monitor.}
\footnotesize
\setlength{\tabcolsep}{3pt}
\begin{tabular}{lccccc}
\toprule
Condition & Live use & Stale effects & File & Git & Network \\
\midrule
PORTICO full & 6/6 & 0/6 & 0/2 & 0/2 & 0/2 \\
Non-revoking comparator & 6/6 & 6/6 & 2/2 & 2/2 & 2/2 \\
All-visible same-policy & 6/6 & 0/6 & 0/2 & 0/2 & 0/2 \\
\bottomrule
\end{tabular}
\vspace{0.2em}
\parbox{0.94\columnwidth}{\footnotesize Family columns report executed stale effects in two cases each.}
\label{tab:capability-live-stale-effects}
\end{table}

\begin{table}[t]
\centering
\caption{Model-live stale-effect follow-up. Static setup materializes a live handle; the model uses it once, then receives a contaminated post-closure reuse instruction.}
\footnotesize
\setlength{\tabcolsep}{3pt}
\begin{tabular}{lccccc}
\toprule
Condition & Live use & Stale effects & File & Git & Network \\
\midrule
PORTICO full & 6/6 & 0/6 & 0/2 & 0/2 & 0/2 \\
Non-revoking comparator & 6/6 & 6/6 & 2/2 & 2/2 & 2/2 \\
All-visible same-policy & 6/6 & 0/6 & 0/2 & 0/2 & 0/2 \\
\bottomrule
\end{tabular}
\vspace{0.2em}
\parbox{0.94\columnwidth}{\footnotesize Family columns report executed stale effects in two cases each.}
\label{tab:capability-live-stale-effects-model}
\end{table}

\paragraph{Action filtering controls a later point.}
Table~\ref{tab:action-filter-visibility} separates two all-visible readings.
The all-visible same-policy comparator keeps broad planner visibility but uses
\system's policy, grants, handle checks, and closure. It matches \system on
forbidden effects (0/20), scope (20/20), and success (5/20), yet logs more
blocked forbidden proposals (4.20 vs.
3.35). Across four post-closure episodes the totals are 84 versus 67; \system records
fewer blocked proposals in three episodes and ties in one
(Appendix Table~\ref{tab:action-filter-visibility-episodes}). The difference is in failures, where broad
visibility blocks 74 proposals versus 57 for \system. This describes planner
interface cost: broader visible request options can still generate extra blocked
actions under the same execution-time policy.

A small boundary-extension check evaluates the same-policy comparator outside
the post-closure family. Both conditions record 0/8 executed forbidden effects and
8/8 scope compliance, but all-visible logs 39 blocked proposals versus 48 for
\system. We therefore keep the post-closure result as a diagnostic of one
family rather than a general trajectory claim.

The coarse all-visible action filter is a deployment diagnostic. It denies stale
uses at invocation time on late-reread rows, but adjacent pressure changes
trajectories and a coarser policy permits effects \system would deny. Over the
full post-grant family it completes 20/20 runs, but only 15/20 remain
scope-compliant; the timeout-boundary detour contributes ten executed forbidden
effects.

\begin{table}[t]
\centering
\caption{Action-filter visibility diagnostic on the post-closure family. The all-visible same-policy condition keeps a broad planner interface but uses \system's execution-time policy; the coarse all-visible action filter keeps broad adjacent authority with coarser checks. Exec. viol. counts executed forbidden effects; Blocked reports average and total denied proposals.}
\scriptsize
\resizebox{\columnwidth}{!}{%
\begin{tabular}{lccccc}
\toprule
Condition & Ep./Runs & Succ. & Scope & Exec. viol. & Blocked avg/total \\
\midrule
\system full & 4/20 & 5/20 & 20/20 & 0/20 & 3.35/67 \\
All-visible same-policy & 4/20 & 5/20 & 20/20 & 0/20 & 4.20/84 \\
Task-wide non-revoking & 4/20 & 5/20 & 20/20 & 0/20 & 3.00/60 \\
Coarse all-visible action filter & 4/20 & 20/20 & 15/20 & 10/20 & 1.25/25 \\
\bottomrule
\end{tabular}
}
\label{tab:action-filter-visibility}
\end{table}

\paragraph{Supporting checks.}
We repeat the closure split across
additional lifecycle and hardening slices (Appendix~\ref{sec:supporting-analyses}). A supplementary frontier slice runs
the grant-lifecycle episodes on GPT-5.5, Gemini 3.5 Flash, and Claude Opus 4.8
under full access, \system, and the non-revoking comparator: GPT-5.5 and Claude
complete 3/3 under \system and no-revoke, Gemini completes 1/3 under both, and
all three preserve scope with zero violations under \system and no-revoke. Full
access completes 3/3 for all profiles, but GPT-5.5 goes out of scope in 2/3
runs. The appendix also covers handle replay, free-path bypass,
resource/effect mismatch, aliases, phase-exit revoke, selector provenance, and
contract-sensitivity replays.

\subsection{RQ4: Lifecycle Behavior Appears on Real Repositories}

\paragraph{Real-repository setup.}
The real-repository suite uses the same contract style on pinned snapshots of
ItsDangerous, Click, Flask, Jinja, Werkzeug, Pluggy, Requests, and HTTPX. It
keeps the lifecycle question task-bounded while replacing the synthetic layout
with real files and module boundaries.

\paragraph{Real-repository results.}
The main Suite-D matrix runs the fourteen real-repository contracts once with
the Qwen3-Coder profile under Full, Static, Sandbox, \system full, the
same-grant no-revoke comparator, and the coarse all-visible action filter. All
six conditions complete 14/14 contracts with 14/14 scope and 0/14 executed
contract-forbidden effects. Only \system full materializes temporary
grants on all 14 contracts. This slice checks that the grant lifecycle and
contract mapping work on real Python project layouts; it is not a broad SWE
performance benchmark.

Supplementary local/API-backed lifecycle roots stress the utility cost of
closure. Across 42 contract-profile cells, \system succeeds in 31/42 with 42/42
scope compliance and 0/42 violations; the whole-task no-revoke diagnostic
completes 42/42. The gap is the cost of bounded closure when contracts close
early, omit a trusted reauthorization path, or depend on a planner that does
not request a fresh grant. The full table is in
Appendix~\ref{sec:supporting-analyses}.

\paragraph{Integration reading.}
A representative Flask run patches \path{src/flask/sessions.py},
receives a declared read-only boundary grant for \path{src/flask/app.py}, and
then denies a packaging detour to \path{pyproject.toml} because it is outside
both the initial envelope and the grant. Detailed real-repository tables are in
Appendix~\ref{sec:supporting-analyses}.

\section{Discussion}
\label{sec:discussion}

\paragraph{What the results mean.}
The key finding is the lifetime split: a task-wide non-revoking comparator can
preserve task success, scope compliance, and unsupported-probe denial before
closure, yet still leave temporary capabilities available. \system removes
that residual exposure. The measured object is the executed effect and the
lifetime that exposed it. The all-visible same-policy comparator adds a narrower
visibility check: broader visibility yields more blocked forbidden proposals on
the post-closure diagnostic without producing executed forbidden effects. A
boundary-extension check reverses the blocked-proposal direction while preserving
safety, so interface pressure remains diagnostic. Closure is the stronger claim;
the coarse all-visible action filter remains a deployment diagnostic with
different policy granularity.

\paragraph{Utility and contracts.}
\system prioritizes scope control over raw success: safe incompletion is
visible, while overbroad success can mask stale authority. Contract quality
remains a deployment input from task and workflow policy; planner
self-attestation cannot create contracts. The timeout sensitivity audit shows
undergranting, overgranting, and wrong-expansion behavior.
Appendix~\ref{sec:supporting-analyses} reports eight deterministic templates
that recover 35/36 structures on held-out transfer; authoring cost remains
separate. The
real-repository experiment shows where deployment policy needs care: an early
closure can block legitimate follow-up work unless the workflow supplies a
trusted reauthorization path. That failure is visible as an incomplete task
rather than a hidden stale capability. Safe recovery should reopen authority
through a trusted event or a narrower request selector. We did not measure
inter-author agreement, review time, or
contract-writing cost.

\paragraph{Revocation scope and external validity.}
Revocation changes the mediated capability interface, not model memory: a
revoked path may remain in context, but the planner cannot invoke it without a
new grant. The guarantee also depends on linearizable mediation: the monitor
must recheck handles before effects, and unmediated shell, network, or file
paths remain outside the claim. Suites A--C are controlled fixtures; Suite D
checks real Python repository layouts. The \texttt{portico-tool} smoke roots
check packaging and public MCP interface behavior, not model performance.
Other build systems, browser tools, SaaS connectors, databases, and notebooks
remain untested. The main model profile is a frozen Hugging Face router trace
root; fresh reruns may select a different backend. Broader deployment needs the
same binding for new tool domains: resource, effect, phase, justification, and
trusted closure.

\section{Related Work}
\label{sec:related-work}

\paragraph*{Adjacent defenses.}
Prompt-injection and misuse benchmarks show that untrusted content can redirect
tool-using agents, including tool-selection hijacking before sandbox
handoff~\cite{promptinjection,injecagent,asb,toolemu,agentdojo,agentharm,agentalign,toolhijacker}.
Guardrails and firewalls sanitize prompts or filter proposed
actions~\cite{promptarmor,llamafirewall,designpatterns,firewalls,causalarmor}.
MCP-security and tool-poisoning research covers descriptor discovery,
provenance, poisoning, and trust propagation~\cite{mcp_landscape,breakingprotocol,securingmcp,smcp,mcpsecbench,mcptox,systematicmcp,mcpitp,ama,maltool}.
\system starts after that layer: it assumes a trusted typed tool interface and
bounds the resource/effect authority exposed by valid descriptors.

\paragraph*{Capability lifetime and revocation.}
Classic least-privilege and capability systems scope, delegate, and revoke
authority~\cite{saltzer,capsicum,cheri}.
Short-lived credentials, leases, epoch invalidation, and usage-control systems
make lifetime part of policy in other settings~\cite{leases,uconabc}. \system applies these mechanisms to the
planner interface of coding agents and measures stale handle capabilities in
trajectories. Its handles connect revocation to model turns,
resource/effect bindings, trusted closure events, and replay checks; the
revoked object is the handle in the next model tool interface. The
mechanism is conventional in spirit, but its
placement is different: the capability is both an execution credential and the
object that shapes the next planner interface.

\paragraph*{Runtime least privilege and isolation.}
Runtime-enforcement and policy-compilation systems, including AgentSpec, PCAS,
Maris, MiniScope, AgentBound, CaMeL, PFI, and Progent, make execution-time
mediation and dynamic policy practical~\cite{agentspec,pcas,maris,miniscope,agentbound,camel,pfi,progent}.
CMTF, RACG, Contract2Tool, and ContractGuard study visible tool menus,
preconditions, effects, and contract integrity~\cite{cmtf,racg,contract2tool,contractguard}.
\system differs in the unit it exposes: a grant is tied to a tool, resource,
effect, phase, and epoch, and closure removes the resulting handle from later
interfaces. Tool-menu filters hide whole primitives, and contract systems decide
whether invocations are allowed; \system applies those ideas to temporary
capabilities whose stale replay is rejected. Isolation and execution-structure
systems constrain reachability or handoff
structure~\cite{isolategpt,sandboxescapebench,agentcgroup,ace,saga,multiagentmemory,cellmate};
\system assumes such mediation and makes a narrower resource/effect lifetime
visible to the planner and enforceable at execution.

\section{Conclusion}
\label{sec:conclusion}

Coding-agent authority has a lifetime. \system makes it explicit in the task
contract: initial capabilities, requestable grants, trusted closure events, and
global deny rules. A boundary request mints opaque resource/effect handles.
Closure removes those handles from the next interface and rejects stale replay
before effects occur.

The experiments isolate that lifetime. A non-revoking comparator receives the
same grant at the same turn, performs the same pre-closure work, and still
permits 10/10 post-closure reuses; \system denies them. Stale-effect audits
cover file writes, git mutation, and network egress, including six live traces.
The same-policy diagnostic keeps forbidden effects at zero while
showing the planning cost of a broader interface. Each grant records its
reason, closure, and renewal path.

\appendix
\section*{Ethical Considerations}

This work studies how to reduce harmful authority in tool-using coding agents.
The artifact includes adversarial scenarios such as dependency mutation,
exfiltration pivots, and unauthorized version-control actions, but they are
intentionally scoped to local benchmark repositories and stubbed resources.
Their purpose is to measure defensive value, not to operationalize misuse
against real systems.

The main stakeholders are repository maintainers, developers using coding
agents, organizations deploying agent runtimes, and researchers who reuse the
artifact. The main potential harms are misuse of the benchmark lures outside a
defensive setting, a false sense of security from an incomplete contract,
overblocking of legitimate developer work, and accidental exposure if someone
adapts the harness to repositories that contain real secrets.

We reduce these risks in four ways. First, the adversarial cases run against
synthetic fixtures or pinned public repositories with no live credentials.
Second, network activity is disabled or routed to stubbed endpoints, and the
benchmark does not include live exfiltration infrastructure. Third, the
artifact is framed as a defensive evaluation harness: it records policy
violations, grant decisions, and revocations rather than providing an
exploitation workflow. Fourth, the paper states the deployment dependency
explicitly: \system is a monitor for explicit task contracts under complete
mediation, not a replacement for sandboxing, review, logging, or careful
contract authoring.

Residual risk remains if a deployment writes an overbroad contract, fails to
mediate a tool path, or treats revocation as model memory erasure. Those cases
fall outside \system's guarantees. A real deployment should combine the monitor
with host isolation, credential management, approval paths for unusual
expansions, and audit review.

\section*{Open Science}
\label{sec:open-science}

An anonymized open-science companion repository for this submission is available
at \url{https://anonymous.4open.science/r/portico/}. It is organized under the
paper title, \emph{Lingering Authority: Revocable Resource-and-Effect
Capabilities for Coding Agents}, and is intended to remain available
throughout anonymous evaluation. The repository is organized as a paper
artifact first: \path{submission_artifact/} contains the frozen result roots,
manifests, independent oracles, generated tables, and reproduction scripts used
by the paper. The standalone public MCP tool lives under \path{portico-tool/}
for users who want to run the request--grant--invoke lifecycle in their own
workspace. The decision-core packages are supporting implementation code rather
than the artifact entry point. This appendix maps the paper's core claims to
the checked-in evidence and points readers to trace-level supporting material.

\subsection*{Artifact and Reproduction Map}
\label{sec:artifact-map}

The companion repository is organized for direct audit of the paper artifact.
The primary reproduction package lives under \path{submission_artifact/}; the
root \path{README.md}, \path{ARTIFACT.md}, \path{REPRODUCING.md}, and
\path{RESULTS.md} explain the layout, frozen roots, regenerated tables, and
short audit commands. The standalone public MCP tool lives under
\path{portico-tool/}. It is useful for trying the request--grant--invoke
lifecycle in a workspace, but the paper results are reproduced from
\path{submission_artifact/}. The decision-core packages under \path{src/portico/}
and \path{submission_artifact/library/} are support code rather than the main
artifact entry point.

Manifest and contract validity can be checked from
\path{submission_artifact/benchmark_v2/manifests/} with
\path{submission_artifact/scripts/validate_benchmark_v2.py}. The main A--C results come
from \path{qwen3_coder_30b_primary_r5_v1},
\path{submission_artifact/benchmark_v2/results/checkpoints.json}, and
\path{generate_benchmark_v2_claim_tables.py}. These files regenerate the main
trade-off and boundary-recovery tables from frozen roots.

Closure and capability-lifetime claims use the post-closure protocol roots
\path{qwen3_coder_30b_b10_capability_protocol_r5_v1},
\path{capability_protocol_local_sanity_v1},
\path{capability_protocol_hardening_v1},
\path{capability_stale_effects_v1},
\path{capability_protocol_overhead_v1}, and the independent oracle files under
\path{submission_artifact/benchmark_v2/oracles/}. The relevant generators are
\path{generate_capability_lifecycle_oracle_audit.py},
\path{generate_capability_stale_effects.py},
\path{generate_capability_protocol_overhead_audit.py}, and
\path{generate_interface_snapshot_table.py}. These roots audit same grant timing,
closure, stale-handle replay, stale write effects, interface snapshots, and
turn/token overhead. The supplementary model-profile lifecycle slice is under
\path{frontier_coding_matrix_handle_fix_v1} and its matching no-\system
baseline is under \path{frontier_coding_matrix_no_portico_v1}.

The public MCP integration is documented in \path{portico-tool/} and
audited through the frozen MCP smoke roots:
\path{portico_mcp_protocol_smoke_v1},
\path{codex_mcp_portico_smoke_v1},
\path{claude_code_mcp_portico_smoke_v1},
\path{opencode_mcp_config_smoke_v1}, and
\path{openhands_mcp_config_smoke_v1}. These smokes record interface snapshot,
authority request, live handle use, closure, and stale replay denial through a
stdio tool interface. These are MCP interface checks rather than
model-evaluation results. The \texttt{portico-tool-smoke} executable and
\path{submission_artifact/Makefile} targets such as \texttt{make mcp-smoke}
provide local checks that require no provider credentials.

Suite D evidence is under \path{submission_artifact/benchmark_v2/real_repo_suite_d_v1/}
and the Qwen3 result roots \path{qwen3_real_repo_boundary_classic_baselines_v1},
\path{qwen3_real_repo_expansion_classic_baselines_v1},
\path{qwen3_real_repo_boundary_lifecycle_comparators_v1}, and
\path{qwen3_real_repo_expansion_lifecycle_comparators_v1}. These roots record
the pinned repository index, real-repository contracts, and grant-bearing cells.
Compound-tool mediation evidence lives under
\path{compound_mediation_audit_v1}. Contract footprint, transfer, and
sensitivity evidence lives under \path{policy_derivation_audit_v1},
\path{policy_derivation_transfer_v2}, and \path{contract_sensitivity_v1}.

The anonymous repository excludes secrets, provider credentials, local virtual
environments, partial runs, and non-essential logs. Upon acceptance, the
artifact can be deanonymized and archived with the same frozen package used
during review.

\bibliographystyle{plain}
\bibliography{refs}

\section{Representative Trace Witnesses}
\label{sec:case-studies}

The main text uses compact running examples. This appendix records the
corresponding checked-in witnesses and the transition each one exercises. Full
JSON traces remain in the artifact.

\paragraph{False detour versus legitimate boundary.}
In the timeout-boundary witness (artifact ID \texttt{B9-02}), the task starts with \path{src/api.py},
\path{tests/test_api.py}, and focused pytest. The contract declares
\path{src/serialization.py} as an on-demand timeout-propagation boundary and
lists \path{notes/packaging.md} as a forbidden side resource. \system denies
the packaging note because no grant rule matches it. A later authority request
mints a serialization handle; the next interface exposes that handle; closure
then removes it. This witness shows that the same primitive can address two
resources while only one has a current justification.

\paragraph{Temporary read closure.}
In the temporary-note witness (artifact ID \texttt{B8-05}), the renderer-repair contract keeps local source, focused
tests, and pytest visible. A recheck note is available only through a temporary
read grant. \system grants the note for inspection, consumes the trusted
subgoal-close event, removes the note handle, and denies a later reread. The
local patch/test handles remain live. This witness shows closure of a temporary
capability without ending the whole repair task.

\paragraph{Same-grant stale handle.}
The stale-handle family gives \system and the non-revoking
comparator the same initial envelope, the same authority request, and the same
grant at the same turn. After closure, \system rejects replay because the
epoch is closed. The comparator keeps the same handle live. This witness is the
minimal trace for isolating capability lifetime.

\section{Protocol Evidence and Supporting Audits}
\label{sec:supporting-analyses}

This appendix keeps evidence that is useful to inspect in the paper itself:
the compiled policy object behind the running example, the lifecycle checks
that exercise stale handles, real-repository evidence, public-tool smoke roots,
and the contract robustness audits. Full per-episode tables, manifest JSON, and
CSV outputs remain in the artifact.

\begin{table}[H]
\centering
\caption{Benchmark suites. Suites A--C are controlled fixtures; Suite D uses pinned real repositories. All contracts are fixed before execution.}
\label{tab:suite-construction}
\scriptsize
\setlength{\tabcolsep}{3pt}
\begin{tabular}{p{0.16\columnwidth} p{0.28\columnwidth} p{0.46\columnwidth}}
\toprule
\textbf{Suite} & \textbf{Runs} & \textbf{Purpose} \\
\midrule
A & 17 scenarios $\times$ 5 repeats per condition & Unsafe exposure with reachable lures and explicit forbidden resources. \\
B & 37 episodes $\times$ 5 repeats per condition & Scoped repair, boundary grants, closure, and same-policy diagnostics. \\
C & 14 episodes $\times$ 5 repeats per condition & Legitimate coding goals surrounded by hostile or overscoped context. \\
D & 14 episodes; 84 Qwen3 real-repository runs; 126 supplementary local/API-backed lifecycle runs & Boundary grants and closure on pinned Python project layouts. \\
\bottomrule
\end{tabular}
\end{table}

\begin{table}[H]
\centering
\caption{Compiled policy excerpt for the timeout-boundary witness. The table shows the part of the contract that matters for the serialization boundary. Planner text does not add authority; only compiled fields and trusted events can change the authority state. The artifact ID is \texttt{B9-02}.}
\scriptsize
\setlength{\tabcolsep}{3pt}
\label{tab:benchmark-protocol}
\label{tab:manifest-fields}
\label{tab:contract-to-policy-map}
\begin{tabular}{p{0.22\columnwidth} p{0.32\columnwidth} p{0.36\columnwidth}}
\toprule
\textbf{Object} & \textbf{Concrete value} & \textbf{Runtime effect} \\
\midrule
Initial handles & API source, API test, focused pytest & Visible at task start. \\
Forbidden resources & \path{notes/packaging.md}, secrets, git push, network egress & Always denied by invariants. \\
Grant rule $\gamma$ & \path{src/serialization.py}; read/write; focused serialization tests & Minted only by the declared boundary request. \\
Closure $\lambda$ & trusted test pass, trusted subgoal close, or phase exit & Removes the grant handles before later turns. \\
Scoring labels & success and violation predicates from the manifest & Used by scorer; not produced by planner rationale. \\
\bottomrule
\end{tabular}
\end{table}

\begin{table}[H]
\centering
\caption{Closure and visibility diagnostics. These checks separate capability lifetime from late execution filtering.}
\scriptsize
\setlength{\tabcolsep}{3pt}
\label{tab:capability-lifecycle-oracle}
\label{tab:closure-lure-summary}
\label{tab:authority-persistence}
\label{tab:action-filter-visibility-episodes}
\label{tab:same-policy-b9-extension}
\begin{tabular}{p{0.25\columnwidth} p{0.31\columnwidth} p{0.34\columnwidth}}
\toprule
\textbf{Diagnostic} & \textbf{Setup} & \textbf{Endpoint} \\
\midrule
Same-grant closure & Same initial envelope, same request, same grant turn. & \system denies 10/10 stale reuses; non-revoking permits 10/10. \\
Same-policy visibility & Same execution policy and closure; broader proposal interface. & Both execute 0/20 forbidden effects; blocked proposals are 67 vs. 84. \\
Boundary-extension visibility & Eight adaptive boundary episodes outside the post-closure family. & Both execute 0/8 forbidden effects; blocked proposals are 48 vs. 39. \\
Frontier lifecycle slice & Three grant-lifecycle episodes across three API-backed model profiles. & The slice compares full access, \system, and non-revoking runs. \\
Local cross-model persistence & Same closure endpoint across local model backbones. & Security direction persists; utility varies by model/runtime. \\
\bottomrule
\end{tabular}
\end{table}

\begin{table}[H]
\centering
\caption{Supplementary API-backed frontier lifecycle slice. The same three grant-lifecycle episodes run under full access, \system, and the non-revoking comparator. This is a narrow protocol check, not the main utility matrix.}
\label{tab:frontier-lifecycle-slice}
\scriptsize
\setlength{\tabcolsep}{3pt}
\begin{tabular}{p{0.22\columnwidth} p{0.27\columnwidth} c c c c}
\toprule
\textbf{Profile} & \textbf{Condition} & \textbf{Success} & \textbf{Scope} & \textbf{Viol.} & \textbf{Grants} \\
\midrule
GPT-5.5 & Full access & 3/3 & 1/3 & 2/3 & -- \\
GPT-5.5 & \system full & 3/3 & 3/3 & 0/3 & 3/3 \\
GPT-5.5 & Non-revoking comparator & 3/3 & 3/3 & 0/3 & 3/3 \\
\midrule
Gemini 3.5 Flash & Full access & 3/3 & 3/3 & 0/3 & -- \\
Gemini 3.5 Flash & \system full & 1/3 & 3/3 & 0/3 & 3/3 \\
Gemini 3.5 Flash & Non-revoking comparator & 1/3 & 3/3 & 0/3 & 3/3 \\
\midrule
Claude Opus 4.8 & Full access & 3/3 & 3/3 & 0/3 & -- \\
Claude Opus 4.8 & \system full & 3/3 & 3/3 & 0/3 & 3/3 \\
Claude Opus 4.8 & Non-revoking comparator & 3/3 & 3/3 & 0/3 & 3/3 \\
\bottomrule
\end{tabular}
\end{table}

\begin{table}[H]
\centering
\caption{Protocol hardening coverage.}
\scriptsize
\setlength{\tabcolsep}{3pt}
\label{tab:controlled-escalation}
\label{tab:capability-protocol-hardening}
\label{tab:capability-protocol-overhead}
\begin{tabular}{p{0.30\columnwidth} p{0.60\columnwidth}}
\toprule
\textbf{Case family} & \textbf{Expected decision pattern} \\
\midrule
Free-path bypass and path aliases & Ambient path attempts are denied unless they resolve to an initial or live handle. \\
Scope and effect mismatch & A handle for one resource/effect cannot authorize a different resource/effect. \\
Phase exit and stale replay & Closure invalidates the epoch; replay is rejected before effects. \\
Selector provenance & Planner-authored rationale selects a candidate rule but cannot create authority. \\
Stale file/git/network effects & \system and all-visible same-policy execute 0/6 stale effects; non-revoking executes 6/6. \\
\bottomrule
\end{tabular}
\end{table}

\begin{table}[H]
\centering
\caption{Real-repository authority-lifecycle evidence. A contract-profile cell is one contract evaluated under one runtime/model profile and aggregated over its fixed executions.}
\label{tab:real-repo-suite-d}
\scriptsize
\setlength{\tabcolsep}{3pt}
\begin{tabular}{p{0.25\columnwidth} p{0.27\columnwidth} p{0.38\columnwidth}}
\toprule
\textbf{Slice} & \textbf{\system endpoint} & \textbf{Comparator reading} \\
\midrule
Qwen3 real-repository matrix, 84 runs & \system full succeeds 14/14 with 14/14 scope, 0/14 violations, and grants on 14/14 contracts. & Full, Static, Sandbox, same-grant no-revoke, and coarse all-visible action filter each succeed 14/14 with 14/14 scope and 0/14 violations. \\
Supplementary Pallets boundary cells, 18 runs & Success 4/6, scope 6/6, violations 0/6. & Whole-task no-revoke succeeds 6/6 without revocation; coarse all-visible succeeds 6/6 with 9 blocked proposals. \\
Supplementary expansion cells, 108 runs & Success 27/36, scope 36/36, violations 0/36. & Whole-task no-revoke succeeds 36/36; coarse all-visible succeeds 35/36 with 42 blocked proposals. \\
\bottomrule
\end{tabular}
\end{table}

\begin{table}[H]
\centering
\caption{Contract derivation and sensitivity. These audits characterize the checked-in contract layer; they do not claim automatic policy synthesis.}
\label{tab:policy-derivation-audit}
\scriptsize
\setlength{\tabcolsep}{3pt}
\begin{tabular}{p{0.28\columnwidth} p{0.30\columnwidth} p{0.32\columnwidth}}
\toprule
\textbf{Audit} & \textbf{Headline result} & \textbf{Reading} \\
\midrule
Template coverage & Eight deterministic templates cover the 70-episode corpus. & Contracts are not per-episode code paths. \\
Holdout transfer & 35/36 paper-subset holdouts match structures derived from non-paper manifests. & Template reuse transfers within the checked-in corpus. \\
Timeout-boundary sensitivity & Removing expansion blocks the boundary; initial visibility removes the grant; wrong-note expansion causes the expected violation. & The monitor inherits contract quality. \\
\bottomrule
\end{tabular}
\end{table}

The omitted appendix material is still part of the artifact: full manifest
excerpts, per-episode post-closure rows, ECES/PTA exposure diagnostics and
weights, overhead CSVs, expanded hardening traces, and generation scripts.
Those files are more useful as auditable records than as compressed PDF tables.

\paragraph{Protocol invariants to inspect.}
The artifact keeps one machine-readable trace per row below. These records are
the most useful checks for a reviewer who wants to inspect the implementation
rather than rerun the whole campaign.
\begin{table}[H]
\centering
\caption{Protocol invariants to inspect in the artifact.}
\scriptsize
\setlength{\tabcolsep}{3pt}
\begin{tabular}{p{0.34\columnwidth} p{0.56\columnwidth}}
\toprule
\textbf{Invariant} & \textbf{Artifact evidence} \\
\midrule
Execution tools consume handles, not ambient paths. & Interface snapshot and free-path bypass traces. \\
Grant IDs and epochs are task-local and checked at execution. & Stale replay, cross-scope, and phase-exit traces. \\
Canonicalization happens before authority comparison. & Alias and normalized-path hardening traces. \\
Closure is driven by trusted observations. & Test-result, orchestrator close, and authenticated revoke traces. \\
Policy perturbations change outcomes predictably. & \texttt{contract\_sensitivity\_v1}. \\
\bottomrule
\end{tabular}
\end{table}

\paragraph{Primary artifact roots.}
The compact roots are \path{qwen3_coder_30b_primary_r5_v1} for the main A--C
campaign, \path{qwen3_coder_30b_b10_capability_protocol_r5_v1} for the
same-grant closure and same-policy diagnostics,
\path{frontier_coding_matrix_handle_fix_v1} and
\path{frontier_coding_matrix_no_portico_v1} for the supplementary API-backed
frontier slice, \path{capability_stale_effects_v1}
and the model-live stale-effect root for post-closure effects, and the
\path{qwen3_real_repo_*} roots for Suite D. The MCP smoke roots audit the
public \texttt{portico-tool} interface. \path{policy_derivation_transfer_v2}
and \path{contract_sensitivity_v1} cover contract robustness.

\paragraph{Utility-cost attribution.}
The main text reports that bounded closure preserves scope but can reduce raw
completion. The failure classes below are the useful diagnostic split: they
separate missing request options from narrow grants, planner instability, and
conservative poisoned-context blocking.
\begin{table}[H]
\centering
\caption{Utility misses on the main repeat-strengthened checkpoint.}
\scriptsize
\setlength{\tabcolsep}{3pt}
\begin{tabular}{p{0.30\columnwidth} p{0.18\columnwidth} p{0.42\columnwidth}}
\toprule
\textbf{Failure class} & \textbf{Runs} & \textbf{Interpretation} \\
\midrule
Missing action option & 15 & Helper-file creation, documentation cross-read, or multi-target patching is not expressed cleanly. \\
Narrow grant insufficient & 20 & The declared expansion stays too small, so runs stop early without measured violations. \\
Planner/model instability & 15 & Dedicated reruns recover several cases, so the miss is not always a policy failure. \\
Conservative poisoned-context blocking & 30 & Zero violations are preserved, but adjacent benign reads or writes can be suppressed. \\
\bottomrule
\end{tabular}
\end{table}

\paragraph{Reauthorization reading.}
The real-repository gap is a contract-lifetime issue, not an unexplained
failure mode. A non-revoking comparator succeeds more often because temporary
authority remains available after the episode closes. A deployable monitor can
recover some of that utility only through a fresh trusted event, such as a new
workflow phase, human approval, or a monitor-observed subgoal transition.
Planner text alone should not reopen a closed epoch.

\begin{table}[H]
\centering
\caption{How to read utility loss under bounded closure.}
\scriptsize
\setlength{\tabcolsep}{3pt}
\begin{tabular}{p{0.28\columnwidth} p{0.31\columnwidth} p{0.31\columnwidth}}
\toprule
\textbf{Cause} & \textbf{Observed symptom} & \textbf{Safe recovery path} \\
\midrule
Closure too early & Work needs the same boundary after a trusted close. & Add a trusted reauthorization event or split the episode. \\
Grant too narrow & Adjacent legitimate files remain denied. & Broaden the declared boundary, not the whole task. \\
Missing request option & Planner cannot ask for the needed boundary. & Add a requestable selector with typed resource/effect bounds. \\
Model instability & Same policy succeeds under a rerun or alternate model. & Treat as planner variance, not a broader authority grant. \\
\bottomrule
\end{tabular}
\end{table}

\paragraph{Protocol cost.}
The two-stage lifecycle adds one materialization turn when a grant is minted
for the next interface. In the post-closure root, \system, the non-revoking
comparator, and the all-visible same-policy comparator all use that turn; the
coarse all-visible action filter does not. The artifact records the associated
step counts, prompt tokens, and harness timestamps.

\paragraph{Mediation boundary.}
The protocol checks assume that runtime calls reach \system before side
effects occur. The \texttt{portico-tool} smokes exercise the stdio interface,
snapshots, authority requests, handle use, closure, and stale replay denial.

\end{document}